\documentclass[journal]{IEEEtran}

\usepackage[dvips]{graphicx}
\usepackage{algorithm}
\usepackage{algpseudocode}
\usepackage{amsfonts}
\usepackage{amsmath}
\usepackage{amssymb}
\usepackage{amsthm}
\usepackage{array}
\usepackage{booktabs}
\usepackage{calc}
\usepackage{cite}
\usepackage{diagbox}
\usepackage{dsfont}
\usepackage{epstopdf,tikz,tikz-network}
\usepackage{graphicx}
\usepackage{hyperref}
\usepackage{multirow}
\usepackage{pgf}
\usepackage{pgfplots}
\usepackage{pgfplotstable}
\usepackage{rotating}
\usepackage{stfloats}
\usepackage{tabularx}
\usepackage{textcomp}
\usepackage{subcaption}
\usepackage{url}
\usepackage{verbatim}
\usepackage{xcolor}
\usetikzlibrary{shapes,arrows}

\begin{document}

	%%%%%%%%%%%%%%%%%%%%%%%%%%%%%%%%%%%%%%%%%%%%%%%%%%%%%%%%%%%%%%%%%%%%%%%%%%%%%%%%
\title{
Joint Communication–Sensing Beamforming and Uplink Power Control for Network-Assisted Full-Duplex Cell-Free ISAC Systems
}

%%%%%%%%%%%%%%%%%%%%%%%%%%%%%%%%%%%%%%%%%%%%%%%%%%%%%%%%%%%%%%%%%%%%%%%%%%%%%%%%
\author{Reza Roshanghias, Seyed Mohammad Karbasi and Reza Saadat 
	\thanks{R.Roshanghias and SM.Karbasi are with the Department of Electrical Engineering, Sharif University of Technology, Tehran, Iran (e-mail: {reza.roshanghias@researcher.sharif.edu},{SM.Karbasi@sharif.edu} ).}
	\thanks{R.Saadat is with the Department of Electrical Engineering Yazd University, Yazd, Iran
		(e-mail: {rsaadat@yazd.ac.ir}).}}

\maketitle

\begin{abstract} In this paper, we investigate the problem of designing joint downlink (DL) communication beamformer and uplink (UL) power as well as sensing beamformer in a multi static integrated sensing and communication (ISAC) enabled cell free massive multiple input multiple output (CF-mMIMO) system. To this goal, we first consider two priority-based beamforming strategies. In the communication-prioritized sensing design, we first design the communication beamformers and then the sensing beamformer is projected onto the nullspace of the effective communication channels to suppress its additional interference to both the DL UEs and UL receivers. In the sensing-prioritized communication design, we select the sensing beamformer first according to the target direction, while the communication beamformers are subsequently optimized using a max--min SINR formulation. In addition to these priority-based designs, we formulate a sensing-centric joint optimization problem  to determine the optimal DL communication beamformers, sensing beamformer, and UL UE transmit powers. Our objective is to maximize the sensing SINR while guaranteeing individual DL and UL SINR requirements and satisfying the per-AP and per-UE transmit-power constraints. Since the resulting problem is non-convex because of the coupled beamforming and UL power variables, the beamforming vectors are lifted into positive-semidefinite transmit covariance matrices and semidefinite relaxation (SDR) is applied. Numerical results demonstrate that the considered framework can simultaneously maintain communication QoS and reliable target sensing under NAFD cross-link interference. The MAPRT detector achieves a detection probability of approximately $0.97$ at $\tau_s=10$, with an empirical false-alarm probability close to $0.05$, while the optimized communication links satisfy their prescribed SINR requirements using the available AP and UE power budgets. \end{abstract}


%=========================================================
% INTRODUCTION
%=========================================================
% NAFD CF-ISAC notation-consistent priority-based beamforming sections

\section{Introduction}
\label{sec:introduction}

The evolution toward sixth-generation (6G) wireless networks is
expected to transform wireless infrastructure from a conventional
communication-only platform into a distributed intelligent network
capable of simultaneously communicating with users and perceiving
the surrounding environment. Integrated sensing and communication
(ISAC) has consequently emerged as an important candidate technology
for future wireless systems because communication and radar sensing
can share spectrum, radio-frequency hardware, antenna arrays, and
signal-processing resources
\cite{Liu2022ISAC,Liu2020JRC,Sturm2011RadarComm,
	Paul2017Survey,Chiriyath2017RadarComm,Wymeersch2017Positioning}.

Comprehensive tutorials and surveys have established the fundamental
principles of ISAC, including waveform design, transmit and receive
beamforming, sensing-assisted communications, communication-assisted
sensing, information-theoretic limits, and radar--communication
performance tradeoffs
\cite{Liu2022ISAC,Liu2020JRC,Paul2017Survey,
	Chiriyath2017RadarComm}. In particular, dual-functional
radar--communication systems exploit a common transmitted waveform
for both data transmission and environmental sensing, thereby
improving spectral and hardware efficiency
\cite{Liu2020JRC,Liu2018MUDFRC}.
Nevertheless, the sensing performance of a conventional centralized
ISAC base station may deteriorate because of blockage, unfavorable
target geometry, severe path loss, and limited spatial diversity.

A complementary architecture for future wireless networks is
cell-free massive MIMO (CF-mMIMO). In contrast to conventional
cellular systems, a large number of geographically distributed APs
cooperatively serve users without relying on fixed cell boundaries
\cite{Ngo2017CF,Nayebi2017CF,Bjornson2020CF}.
The distributed architecture provides macro-diversity and can reduce
the large service-quality variations associated with cell-edge users.
The seminal work in \cite{Ngo2017CF} demonstrated the substantial
performance advantages of cell-free massive MIMO over conventional
small-cell deployments. Subsequent studies investigated power
control, precoding, channel estimation, centralized and distributed
processing, energy efficiency, and scalable implementations
\cite{Nayebi2017CF,Ngo2018Energy,Bjornson2020CF,
	Shaik2020RadioStripes}.

The theoretical and practical foundations of user-centric cell-free
massive MIMO have been systematically developed in
\cite{Demir2021Foundations}, while comprehensive surveys such as
\cite{Elhoushy2022Survey,Ammar2022Survey} discuss channel estimation,
pilot assignment, user association, power allocation, fronthaul
requirements, and scalable cooperation. These studies establish
CF-mMIMO as a flexible distributed architecture in which each UE can
be jointly served by a suitable subset of geographically distributed
APs.

The distributed nature of cell-free networks is particularly
attractive for sensing. Instead of observing a target from a single
location, spatially separated APs can illuminate and observe the
environment from different directions, creating a multi-static
sensing architecture with improved spatial diversity. This
observation has motivated the emerging concept of cell-free ISAC
(CF-ISAC)
\cite{Galappaththige2025CFISAC,CFISACFronthaul2024,
	Elbir2024Hybrid,Nguyen2024Multistatic}.
A recent comprehensive treatment of CF-ISAC
\cite{Galappaththige2025CFISAC} discusses cooperative transmission,
multi-static sensing, target-centric architectures, sensing metrics,
resource allocation, synchronization, interference management, and
fronthaul limitations.

Several recent works have investigated different aspects of CF-ISAC.
For example, capacity-constrained fronthaul and hybrid precoding were
studied in \cite{CFISACFronthaul2024}, while hybrid beamforming for
distributed ISAC was considered in \cite{Elbir2024Hybrid}.
Multi-static CF-mMIMO ISAC performance and power allocation were
investigated in \cite{Nguyen2024Multistatic}. Dynamic AP operation
mode selection for mmWave CF-ISAC was considered in
\cite{Yan2024ModeSelection}, highlighting the importance of deciding
whether individual APs should communicate, sense, or remain inactive.

Resource allocation and beamforming have also received increasing
attention. Joint communication and sensing beamforming under security
requirements was studied in \cite{Ren2024Secure}, while joint secrecy
rate and sensing-SNR optimization was investigated in
\cite{SecureCFISAC2024}. Learning-based distributed beamforming
approaches have been proposed to reduce the computational and
fronthaul requirements of centralized optimization
\cite{Elrashidy2025Learning,Masood2024Beamforming}.
More recently, joint space-time adaptive processing and beamforming
has been considered for cooperative CF-ISAC target detection
\cite{Liu2025STAP}, and graph-learning approaches have been developed
to bridge system-level CF-ISAC optimization and target-parameter
estimation \cite{GraphCFISAC2026}.

Target detection constitutes another important aspect of practical
CF-ISAC operation. Many resource-allocation studies implicitly assume
that the target exists and that its relevant parameters are known.
In practice, however, the network must first determine whether a
target is present before allocating communication and sensing
resources based on target-dependent channels. Recent work has
therefore begun explicitly incorporating target detection into
cell-free ISAC. In \cite{Singh2025OTFS}, an OTFS-aided cell-free
MIMO ISAC system was investigated and a maximum a posteriori ratio
test was developed for target detection. Multiple-target detection
in CF-mMIMO-assisted ISAC was studied in
\cite{Elfiatoure2025MultiTarget}, demonstrating the importance of
distributed AP cooperation for sensing.

In parallel with these developments, full-duplex communication has
long been recognized as a means of improving spectral efficiency by
supporting simultaneous transmission and reception in the same
frequency band
\cite{Sabharwal2014FD,Bharadia2013FD,Zhang2016FD}.
Its practical performance, however, is strongly affected by
self-interference and cross-link interference. These challenges
become even more significant in ISAC systems because strong
communication signals can mask weak target echoes
\cite{Tang2024FDISAC,Jiang2024FDISAC}.

Network-assisted full-duplex (NAFD) provides an alternative approach
to conventional full-duplex radios. Instead of requiring each AP to
transmit and receive simultaneously, geographically distributed
half-duplex APs are divided into DL and UL groups. The DL APs and UL
APs simultaneously serve DL and UL users over the same
time-frequency resources, thereby realizing full-duplex operation at
the network level
\cite{Wang2019NAFD,Mohammadi2023NAFD,Li2023NAFD,Okui2023NAFD}.
This architecture avoids strong self-interference at each individual
AP but introduces inter-AP interference from DL APs to UL APs and
UL-to-DL cross-link interference between users.

The performance of NAFD CF-mMIMO under imperfect CSI was investigated
in \cite{Wang2019NAFD}. Joint AP-mode assignment, power control, and
large-scale fading decoding for spectral- and energy-efficiency
optimization were subsequently developed in
\cite{Mohammadi2023NAFD}. Limited fronthaul and quantized
implementations were investigated in \cite{Li2023NAFD,Okui2023NAFD}.
These studies demonstrate the potential of NAFD cell-free networks
for simultaneous UL and DL communications, but they primarily focus
on communication performance rather than the coexistence of
communication, sensing, target detection, and target-dependent
inter-AP interference.

The combination of NAFD and CF-ISAC creates a particularly challenging
interference environment. During simultaneous transmission, a UL AP
receives the desired UL signals together with interference generated
by the DL APs. When a sensing target is present, the DL waveform also
produces target-reflected components that are observed at the UL APs.
At the same time, UL UE transmissions interfere with DL reception,
while communication and sensing beams compete for the limited
transmit-power budgets of the DL APs. Hence, communication
beamforming, sensing beamforming, UL power allocation, channel
estimation, and target detection become strongly coupled.

Although SDR, successive convex approximation, alternating
optimization, and related optimization techniques have been widely
used for non-convex beamforming and resource-allocation problems
\cite{Luo2010SDR,Liu2018MUDFRC,Tang2024FDISAC,
	Nguyen2024Multistatic}, the joint treatment of these mechanisms in an
NAFD cell-free ISAC architecture remains comparatively unexplored.
In particular, assuming perfect knowledge of target-dependent
channels before establishing target existence may result in an
optimistic system model.

Motivated by these observations, this work considers an NAFD
cell-free ISAC network in which distributed APs are partitioned into
DL and UL operating sets. The DL APs jointly transmit communication
and sensing signals, while the UL APs simultaneously receive UL
communication signals and target-related sensing components. Channel
estimation and target detection are explicitly incorporated into the
transmission protocol, and a MAPRT detector is employed before
target-dependent optimization is performed.

%=========================================================
% CONTRIBUTIONS
%=========================================================

\subsection{Main Contributions}

The main contributions of this paper are summarized as follows.

\begin{itemize}
	
	\item
	We develop an interference-aware NAFD cell-free ISAC framework that
	jointly supports DL communication, UL communication, and distributed
	target sensing. The considered model explicitly incorporates imperfect
	channel estimation, residual DL-to-UL cross-link interference,
	UL-to-DL interference, and target-induced sensing links. A four-stage
	transmission protocol is adopted, consisting of channel estimation,
	MAPRT-based target detection, target-assisted inter-AP channel
	estimation, and simultaneous communication--sensing transmission.
	
	\item
	We formulate a sensing-centric joint optimization problem for the DL
	communication beamformers, sensing beamformer, and UL UE transmit
	powers. The objective is to maximize the sensing SINR while satisfying
	the prescribed DL and UL communication SINR requirements, per-AP
	transmit-power constraints, and UL UE power limits. This formulation
	explicitly captures the coupling between sensing illumination,
	communication beamforming, UL power control, and NAFD cross-link
	interference.
	
	\item
	To address the resulting non-convex optimization problem, we transform
	the beamforming vectors into positive-semidefinite covariance matrices
	and employ semidefinite relaxation (SDR). An alternating-optimization
	(AO) algorithm is developed to alternately optimize the DL
	communication/sensing covariance matrices and the UL transmit powers.
	For fixed UL powers, the beamforming subproblem is handled through SDR
	and bisection over the achievable sensing SINR, whereas the UL power
	variables are subsequently updated for the obtained beamforming
	solution. This procedure is repeated until convergence.
	
	\item
	In addition to the joint optimization framework, we develop two
	complementary priority-based beamforming strategies to characterize
	the fundamental communication--sensing tradeoff. In the
	\emph{communication-prioritized sensing (CPS)} strategy, the
	communication beamformers are designed first, while the sensing
	beamformer is projected onto the nullspace of the effective
	communication channels. Unlike conventional nullspace sensing designs,
	the proposed construction accounts for both the DL UE channels and
	the effective DL-to-UL cross-link channels after UL receive combining,
	thereby protecting both DL and UL communication from additional
	sensing-induced interference.
	
	\item
	In the \emph{sensing-prioritized communication (SPC)} strategy, the
	sensing beamformer is designed first according to the dominant target
	sensing direction and is subsequently kept fixed. The communication
	beamformers are then optimized through a max--min SINR formulation
	subject to the NAFD interference structure and per-AP power
	constraints. Hence, the CPS and SPC schemes represent two opposite
	operating points: the former prioritizes communication protection and
	uses the remaining spatial degrees of freedom for sensing, whereas the
	latter prioritizes target illumination and utilizes the remaining
	degrees of freedom to improve communication performance.
	
	\item
	Target detection is explicitly incorporated through a
	maximum-a-posteriori ratio test (MAPRT), allowing the sensing
	performance to be evaluated in terms of detection, false-alarm, and
	misdetection probabilities rather than relying solely on sensing SINR.
	The numerical results demonstrate the interaction among target
	detection performance, sensing SINR, communication QoS, UL power
	allocation, and NAFD cross-link interference, and enable a direct
	comparison between the proposed joint design and the two
	priority-based beamforming strategies.
	
\end{itemize}

\textbf{To the best of our knowledge, the main novelty of this work is
	the unified treatment of MAPRT-based target detection, imperfect
	channel knowledge, NAFD cross-link interference, distributed
	communication/sensing beamforming, and UL power control within a
	single SDR--AO optimization framework for cell-free ISAC.}
In contrast to conventional CF-ISAC formulations that assume the
sensing target is already known to exist, the proposed framework
explicitly connects the target-detection stage to the subsequent
target-aware interference-management and resource-allocation stage.

%=========================================================
% PAPER ORGANIZATION
%=========================================================

The remainder of this paper is organized as follows.
Section~II presents the NAFD cell-free ISAC system model, channel
models, channel-estimation procedure, and the DL, UL, and sensing
signal models. Section~III describes the target-detection stage and
the MAPRT detector and formulates the joint communication--sensing
optimization problem in vector form. Section~IV transforms the
problem into its matrix SDR representation and develops the proposed
AO solution. Section~V presents the numerical results, and
Section~VI concludes the paper.

%=========================================================
\section{System Configuration and Notation}
%=========================================================

Consider an integrated sensing and communication (ISAC)-enabled 
cell-free massive MIMO (CF-mMIMO) system operating over a region 
traversed by a mobile target. A central processing unit (CPU) coordinates 
$M$ geographically distributed half-duplex (HD) access points (APs), 
each equipped with $N$ antennas, to simultaneously provide communication 
services to $K$ single-antenna user equipments (UEs) and sense a 
point-like device-free target over the same time-frequency resources. 
Under CPU-based scheduling, the APs are partitioned into two disjoint 
groups. Specifically, $M_{\rm UL}$ APs operate in the uplink (UL) 
reception mode, whereas the remaining $M_{\rm DL}$ APs operate in the 
downlink (DL) transmission mode, where
\begin{equation}
	M_{\rm UL}+M_{\rm DL}=M.
\end{equation}
The corresponding AP index sets are denoted by 
$\mathcal{M}_{\rm UL}$ and $\mathcal{M}_{\rm DL}$, respectively, such that
\begin{equation}
	\mathcal{M}_{\rm UL}\cap\mathcal{M}_{\rm DL}=\varnothing,
	\qquad
	\mathcal{M}_{\rm UL}\cup\mathcal{M}_{\rm DL}=\mathcal{M}.
\end{equation}

\begin{figure}[t]
	\centering
	\includegraphics[width=\linewidth]{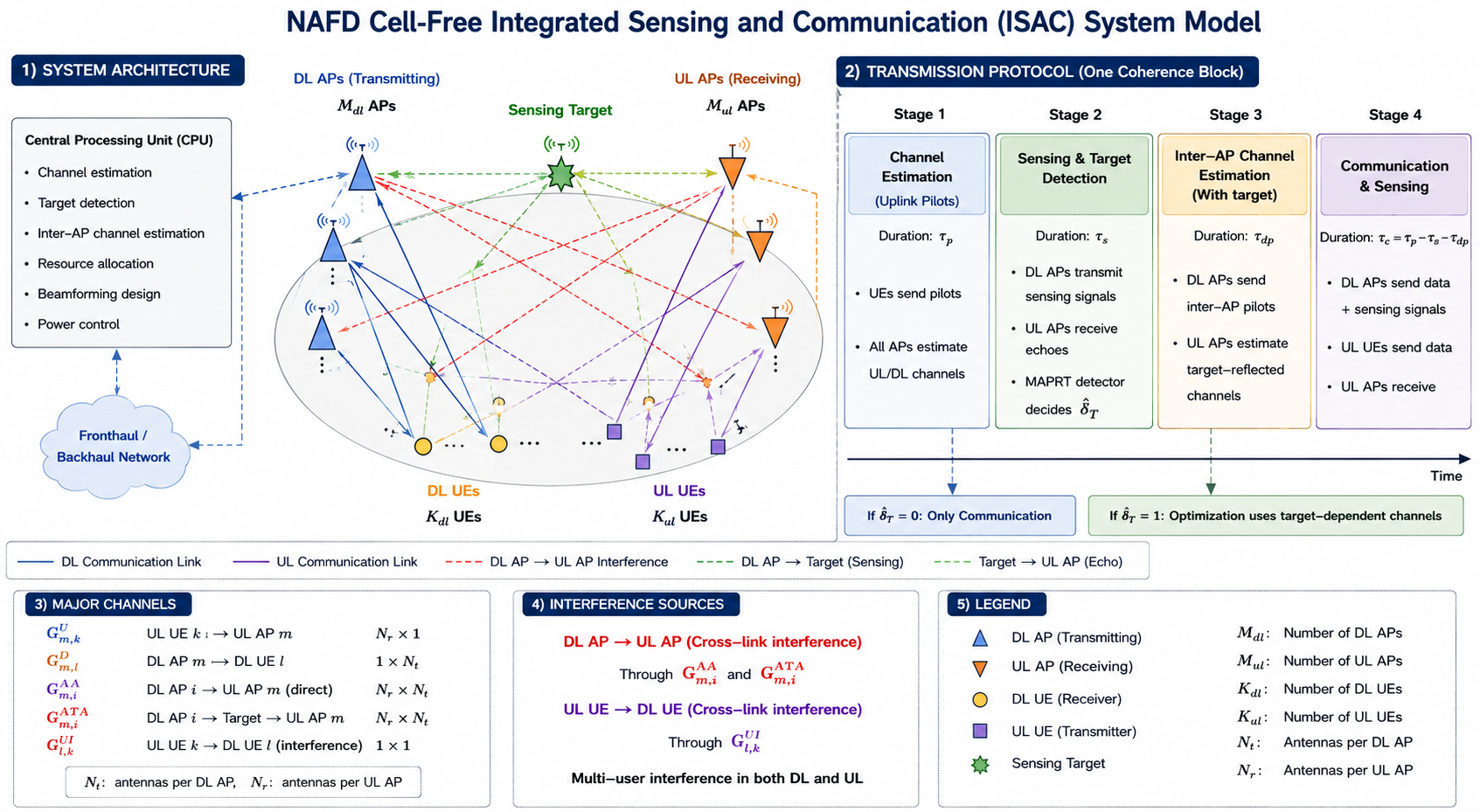}
	\caption{Our Proposed System Model}
	\label{fig:system_model}
\end{figure}

Similarly, among the $K$ UEs, $K_{\rm UL}$ UEs request UL transmission 
and the remaining $K_{\rm DL}$ UEs request DL service, with
\begin{equation}
	K_{\rm UL}+K_{\rm DL}=K.
\end{equation}
Their corresponding index sets are denoted by 
$\mathcal{K}_{\rm UL}$ and $\mathcal{K}_{\rm DL}$, respectively, where
\begin{equation}
	\mathcal{K}_{\rm UL}\cap\mathcal{K}_{\rm DL}=\varnothing,
	\qquad
	\mathcal{K}_{\rm UL}\cup\mathcal{K}_{\rm DL}=\mathcal{K}.
\end{equation}
Figure~\ref{fig:system_model} illustrates the considered NAFD cell-free ISAC architecture and its transmission protocol. The distributed APs are divided into DL and UL sets, where the DL APs simultaneously serve DL UEs and illuminate the sensing target, while the UL APs receive UL transmissions and target echoes. The system operates over four stages comprising channel estimation, MAPRT-based target detection, target-aware inter-AP channel estimation, and joint communication--sensing transmission. The figure also highlights the major propagation channels and cross-link interference, particularly DL-AP-to-UL-AP and UL-UE-to-DL-UE interference, which motivate the proposed joint beamforming and power-control design.

The effective inter-AP channel consists of a target-free propagation 
component and, when the target is present, a target-reflected component. 
Consequently, the target presence must be determined before estimating 
the inter-AP channels. To this end, a dedicated target-detection stage 
is performed, during which all DL APs transmit sensing signals over 
$\tau_s$ symbols, while all UL APs cooperatively process the received 
echoes to determine the estimated target-presence indicator 
$\widehat{\delta}_T$. We define the actual target indicator as
\begin{equation}
	\delta_T \in \{0,1\},
\end{equation}
where $\delta_T=1$ indicates target presence and $\delta_T=0$ indicates 
target absence.

A quasi-static block-fading model is adopted, where each coherence 
interval comprises $\tau_c$ symbols. Exploiting the temporal 
continuity of the target motion, the CPU uses the sensing result obtained 
in the preceding coherence interval to acquire prior information about 
the target position, denoted by
\begin{equation}
	(\bar{x}_T,\bar{y}_T),
\end{equation}
. A hotspot region centered at 
$(\bar{x}_T,\bar{y}_T)$ is subsequently defined to represent the region 
in which the target is most likely to appear during the current coherence 
interval.

When $\delta_T=1$, the target is assumed to lie within this hotspot 
region, and the system performs target sensing, including the estimation 
of its motion characteristics such as velocity and direction, concurrently 
with UL and DL communications. In contrast, when $\delta_T=0$, no target 
is present within the considered sensing region and the available resources 
are devoted solely to communication. This assumption is motivated by the 
continuity of target motion over consecutive coherence intervals.
\subsection{Stage Arrangement of the Proposed Interference Management Mechanism}

The proposed interference management mechanism incorporates both direct
and indirect interference suppression techniques. Each coherence interval,
consisting of $\tau_c$ symbols, is divided into four consecutive stages,
as illustrated in Fig.~\ref{fig:system_model}.

\subsubsection{Uplink Pilot Training Stage}

During the first stage, all UEs transmit uplink pilot sequences of length
$\tau_p^{\rm UL}$. Based on the received pilot signals, the CPU estimates
the channel state information (CSI) between the UEs and the APs. The
resulting channel estimates are subsequently employed for UL signal
reception and DL transmission.

\subsubsection{Target Detection Stage}

The inter-AP channel between a DL AP and a UL AP depends on the presence
of the target. When the target is absent, i.e., $\delta_T=0$, the inter-AP
channel contains only the target-free propagation component. In contrast,
when the target is present, i.e., $\delta_T=1$, the effective inter-AP
channel consists of both the target-free component and the target-reflected
component. Therefore, determining the target presence prior to inter-AP
channel estimation is essential.

During this stage, all DL APs transmit sensing signals of length $\tau_s$,
while the UL APs cooperatively process the received signals to detect the
target. The resulting target-presence estimate is denoted by
$\widehat{\delta}_T$.

\subsubsection{Downlink Pilot Training Stage}

During the third stage, all DL APs transmit downlink pilot sequences of
length $\tau_p^{\rm DL}$, which are jointly received by the UL APs.
Using these received pilot signals together with the estimated target
indicator $\widehat{\delta}_T$, the CPU estimates the effective inter-AP
CSI. In particular, the target-detection result enables the CPU to account
for the target-reflected propagation component when the target is detected.

\subsubsection{Data Transmission and Target Sensing Stage}

The remaining symbols of each coherence interval are allocated to
simultaneous UL/DL data transmission and, when required, target sensing.
Accordingly, the duration of the fourth stage is
\begin{equation}
	\tau_d =
	\tau_c-\tau_p^{\rm UL}-\tau_s-\tau_p^{\rm DL}.
\end{equation}

The sensing operation is activated according to the estimated
target-presence indicator $\widehat{\delta}_T$. When
$\widehat{\delta}_T=1$, communication and target sensing are performed
concurrently over the same time-frequency resources. In this case, the
CPU jointly coordinates the DL APs and UL APs to support UL/DL
communications while sensing the target and suppressing the associated
communication, sensing, and cross-link interference. In contrast, when
$\widehat{\delta}_T=0$, no target is detected and the system operates in
a communication-only mode.

Therefore, the operation of the fourth stage can be summarized as
\begin{equation}
	\widehat{\delta}_T =
	\begin{cases}
		1, & \text{UL/DL communication and target sensing},\\
		0, & \text{UL/DL communication only}.
	\end{cases}
\end{equation}

\begin{figure}[t]
	\centering
	\includegraphics[width=\linewidth]{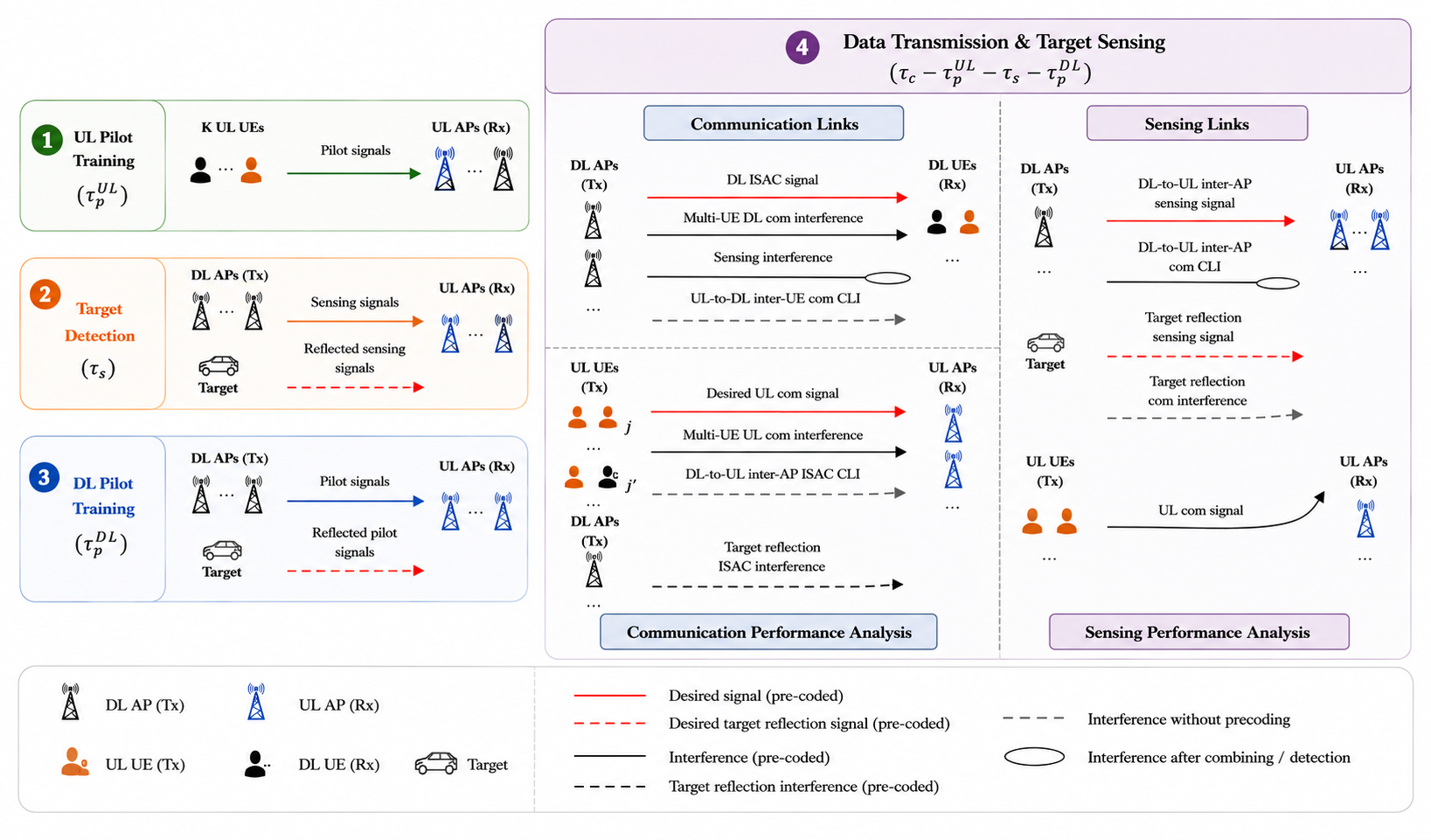}
	\caption{Our Proposed System Model}
	\label{fig:Interference Management}
\end{figure}

\subsection{Channel Model} \label{sec:channel_model}

We consider a quasi-static block-fading model in which all wireless
channels remain approximately constant within each coherence interval
and vary independently between different coherence intervals. Moreover,
the channels are assumed to be frequency-flat over the considered
bandwidth.

Let $\mathbf{h}_{m,k}^{\rm UA}\in\mathbb{C}^{N\times 1}$ denote the
channel between UE $k$ and AP $m$. In contrast to explicitly decomposing
the channel into large-scale and small-scale fading components, we
incorporate the effects of propagation loss, shadowing, and small-scale
fading directly into $\mathbf{h}_{m,k}^{\rm UA}$. Accordingly, the
channel is modeled as
\begin{equation}
	\mathbf{h}_{m,k}^{\rm UA}
	\sim
	\mathcal{CN}
	\left(
	\mathbf{0},
	\mathbf{R}_{m,k}^{\rm UA}
	\right),
\end{equation}
where $\mathbf{R}_{m,k}^{\rm UA}\in\mathbb{C}^{N\times N}$ denotes the
spatial channel covariance matrix and implicitly accounts for the
average channel power and the corresponding large-scale propagation
effects. For spatially uncorrelated Rayleigh fading, this model reduces
to
\begin{equation}
	\mathbf{R}_{m,k}^{\rm UA}
	=
	\beta_{m,k}^{\rm UA}\mathbf{I}_{N},
\end{equation}
where $\beta_{m,k}^{\rm UA}$ represents the average channel power.

Similarly, the cross-link interference channel from UL UE $j$ to DL
UE $l$ is denoted by
\begin{equation}
	h_{j,l}^{\rm UU}
	\sim
	\mathcal{CN}
	\left(
	0,
	\beta_{j,l}^{\rm UU}
	\right),
\end{equation}
where $\beta_{j,l}^{\rm UU}$ implicitly captures the corresponding
large-scale propagation effects.

The target-free channel between DL AP $i$ and UL AP $m$ is denoted by
$\mathbf{H}_{m,i}^{\rm AA}\in\mathbb{C}^{N\times N}$ and modeled as
\begin{equation}
	\mathbf{H}_{m,i}^{\rm AA}
	\sim
	\mathcal{CN}
	\left(
	\mathbf{0},
	\mathbf{R}_{m,i}^{\rm AA}
	\right),
\end{equation}
where $\mathbf{R}_{m,i}^{\rm AA}$ incorporates the average propagation
gain as well as the spatial characteristics of the inter-AP channel.
Hence, no separate path-loss coefficient or AP-to-AP distance variable
is required in the subsequent formulation.

\subsubsection{Target-Reflected Channel}

Since the considered device-free target is a mobile outdoor object,
such as a vehicle or an unmanned aerial vehicle (UAV), the propagation
links between the APs and the target are assumed to be dominated by
line-of-sight (LoS) components. Furthermore, all APs are equipped with
uniform linear arrays (ULAs) with half-wavelength antenna spacing.

Let
\begin{equation}
	\mathbf{a}(\theta)
	\triangleq
	\frac{1}{\sqrt{N}}
	\left[
	1,
	e^{j\pi\sin(\theta)},
	\ldots,
	e^{j\pi(N-1)\sin(\theta)}
	\right]^{T}
\end{equation}
denote the normalized ULA steering vector, where $\theta$ represents
the corresponding angular direction.

Let $\theta_{dl,i}$ denote the direction from DL AP $i$ toward the
target, and let $\theta_{ul,m}$ denote the direction from the target
toward UL AP $m$. Following the bistatic sensing model, the
target-reflected channel from DL AP $i$ to UL AP $m$ through the target
is expressed as
\begin{equation}
	\boxed{
		\mathbf{H}_{m,i}^{\rm ATA}
		=
		\vartheta_{m,i}^{\rm RCS}
		\lambda_{m,i}^{\rm ATA}
		\mathbf{a}(\theta_{ul,m})
		\mathbf{a}^{H}(\theta_{dl,i})
	}
	\label{eq:ATA_channel}
\end{equation}
where $\vartheta_{m,i}^{\rm RCS}$ denotes the bistatic radar
cross-section (RCS) of the target associated with the reflection path
from DL AP $i$ to UL AP $m$.

Following the Swerling-I target model, the target is assumed to move
slowly relative to the sensing duration. Therefore,
$\vartheta_{m,i}^{\rm RCS}$ remains approximately constant over the
sensing symbols collected within consecutive coherence intervals.
Since the RCS depends on the physical characteristics of the target,
including its shape, material composition, and the incidence angle of
the illuminating waveform, the CPU is assumed to possess an estimate
$\widehat{\vartheta}_{m,i}^{\rm RCS}$ obtained from prior sensing
observations.

Following the target-reflected propagation model, the coefficient in
\eqref{eq:ATA_channel} is given by
\begin{equation}
	\lambda_{m,i}^{\rm ATA}
	=
	\frac{\lambda_c}
	{\sqrt{
			(4\pi)^3
			d_{{\rm TA},dl,i}^{\,2}
			d_{{\rm TA},ul,m}^{\,2}
	}},
\end{equation}
where $\lambda_c$ denotes the carrier wavelength,
$d_{{\rm TA},dl,i}$ is the distance between DL AP $i$ and the target,
and $d_{{\rm TA},ul,m}$ is the distance between the target and UL AP
$m$.

Using the prior target-position estimate $(\bar{x}_T,\bar{y}_T)$, the
CPU evaluates the nominal propagation geometry of the reflected path.
The uncertainty associated with the target location and RCS estimation
is incorporated into the statistical model of
$\mathbf{H}_{m,i}^{\rm ATA}$.

Finally, the effective channel between DL AP $i$ and UL AP $m$ can be
compactly represented as
\begin{equation}
	\boxed{
		\mathbf{H}_{m,i}^{\rm eff}
		=
		\mathbf{H}_{m,i}^{\rm AA}
		+
		\delta_T
		\mathbf{H}_{m,i}^{\rm ATA}
	}
	\label{eq:effective_interAP_channel}
\end{equation}
where $\delta_T\in\{0,1\}$ is the target-presence indicator. Therefore,
when $\delta_T=0$, the effective inter-AP channel contains only the
target-free component, whereas for $\delta_T=1$, it consists of both
the target-free and target-reflected propagation components.
\section{Channel Estimation in the Uplink Pilot Training Stage}

During the uplink pilot training stage, the $K$ UEs transmit pilot
sequences for channel estimation. Let
\[
\boldsymbol{\phi}_{1}^{\rm p},
\ldots,
\boldsymbol{\phi}_{\tau_p}^{\rm p}
\in \mathbb{C}^{\tau_p}
\]
denote a set of mutually orthogonal pilot sequences satisfying

\begin{equation}
	\left(\boldsymbol{\phi}_{t}^{\rm p}\right)^H
	\boldsymbol{\phi}_{t'}^{\rm p}
	=
	\begin{cases}
		\tau_p, & t=t',\\
		0, & t\neq t'.
	\end{cases}
\end{equation}

Let $t_k\in\{1,\ldots,\tau_p\}$ denote the pilot assigned to UE $k$.
The set of UEs sharing the same pilot as UE $k$ is defined as

\begin{equation}
	\mathcal P_{t_k}
	=
	\{\, i \; | \; t_i=t_k \,\}.
\end{equation}

The received pilot signal at AP $m$ is given by

\begin{equation}
	\mathbf Y_m^{\rm p}
	=
	\sum_{k=1}^{K}
	\sqrt{\tau_p \rho_p}\,
	\mathbf h_{m,k}^{\rm UA}
	\left(\boldsymbol{\phi}_{t_k}^{\rm p}\right)^H
	+
	\mathbf N_m^{\rm p},
	\label{eq:Ypilot}
\end{equation}

where $\rho_p$ denotes the pilot transmit power and

\[
\mathbf N_m^{\rm p}
\in
\mathbb C^{N\times \tau_p}
\]

contains independent complex Gaussian noise samples with variance
$\sigma_{\rm ul}^2$.

To estimate the channel between UE $k$ and AP $m$, the received pilot
matrix is correlated with the corresponding pilot sequence,

\begin{equation}
	\mathbf y_{m,k}^{\rm p}
	=
	\mathbf Y_m^{\rm p}
	\boldsymbol{\phi}_{t_k}^{\rm p}
	=
	\sqrt{\tau_p\rho_p}
	\mathbf h_{m,k}^{\rm UA}
	+
	\sum_{i\in\mathcal P_{t_k}\setminus\{k\}}
	\sqrt{\tau_p\rho_p}
	\mathbf h_{m,i}^{\rm UA}
	+
	\mathbf N_m^{\rm p}
	\boldsymbol{\phi}_{t_k}^{\rm p}.
	\label{eq:pilot_projection}
\end{equation}

Following the MMSE channel estimation framework, the channel
$\mathbf h_{m,k}^{\rm UA}$ is modeled as

\begin{equation}
	\mathbf h_{m,k}^{\rm UA}
	\sim
	\mathcal{CN}
	\!\left(
	\mathbf 0,
	\mathbf R_{m,k}^{\rm UA}
	\right),
\end{equation}

where $\mathbf R_{m,k}^{\rm UA}$ denotes the channel covariance matrix.

The MMSE estimate of
$\mathbf h_{m,k}^{\rm UA}$
is

\begin{equation}
	\widehat{\mathbf h}_{m,k}^{\rm UA}
	=
	\sqrt{\tau_p\rho_p}
	\mathbf R_{m,k}^{\rm UA}
	\mathbf\Psi_{t_k,m}^{-1}
	\mathbf y_{m,k}^{\rm p},
	\label{eq:mmse_estimator}
\end{equation}

where

\begin{equation}
	\mathbf\Psi_{t_k,m}
	=
	\sum_{i\in\mathcal P_{t_k}}
	\tau_p\rho_p
	\mathbf R_{m,i}^{\rm UA}
	+
	\sigma_{\rm ul}^{2}
	\mathbf I_N.
	\label{eq:Psi}
\end{equation}

The estimated channel is distributed as

\begin{equation}
	\widehat{\mathbf h}_{m,k}^{\rm UA}
	\sim
	\mathcal{CN}
	\left(
	\mathbf 0,
	\mathbf\Phi_{m,k}
	\right),
\end{equation}

with covariance matrix

\begin{equation}
	\mathbf\Phi_{m,k}
	=
	\tau_p\rho_p
	\mathbf R_{m,k}^{\rm UA}
	\mathbf\Psi_{t_k,m}^{-1}
	\mathbf R_{m,k}^{\rm UA}.
	\label{eq:Phi}
\end{equation}

By the orthogonality property of MMSE estimation,
the true channel can be decomposed as

\begin{equation}
	\mathbf h_{m,k}^{\rm UA}
	=
	\widehat{\mathbf h}_{m,k}^{\rm UA}
	+
	\widetilde{\mathbf h}_{m,k}^{\rm UA},
	\label{eq:channel_decomp}
\end{equation}

where the estimation error
$\widetilde{\mathbf h}_{m,k}^{\rm UA}$
is statistically independent of
$\widehat{\mathbf h}_{m,k}^{\rm UA}$ and follows

\begin{equation}
	\widetilde{\mathbf h}_{m,k}^{\rm UA}
	\sim
	\mathcal{CN}
	\left(
	\mathbf 0,
	\mathbf C_{m,k}
	\right),
\end{equation}

with error covariance matrix

\begin{equation}
	\mathbf C_{m,k}
	=
	\mathbf R_{m,k}^{\rm UA}
	-
	\mathbf\Phi_{m,k}.
	\label{eq:Cmk}
\end{equation}
\section{Target Detection Stage}
\label{sec:target_detection}

During the target detection stage, all DL APs transmit dedicated and
mutually independent sensing waveforms, while the UL APs cooperatively
process the received sensing signals to determine whether the target is
present within the hotspot region. Since the primary objective of this
stage is target-presence detection, the transmitted waveforms are
designed to enhance the sensing performance.

For DL AP $i\in\mathcal{M}_{\rm DL}$, the transmitted sensing signal
over $\tau_s$ sensing symbols is expressed as
\begin{equation}
	\mathbf{x}_i^{s}[t]
	=
	\mathbf{w}_i^{s}s_i^{s}[t],
	\qquad t=1,\ldots,\tau_s,
\end{equation}
where $s_i^{s}[t]$ denotes the $t$-th symbol of the dedicated sensing
sequence $\mathbf{s}_i^{s}\in\mathbb{C}^{\tau_s}$, which is normalized
to unit power, and $\mathbf{w}_i^{s}\in\mathbb{C}^{N\times 1}$ denotes
the sensing precoding vector.

Under the single-target sensing model, conjugate beamforming toward the
predicted target location provides an effective sensing precoder since
it coherently concentrates the transmitted power toward the target.
Accordingly, the sensing precoder at DL AP $i$ is selected as
\begin{equation}
	\mathbf{w}_i^{s}
	=
	\sqrt{P_{\rm AP}}\,
	\mathbf{w}_i^{\rm cp},
\end{equation}
where
\begin{equation}
	\mathbf{w}_i^{\rm cp}
	=
	\frac{
		\mathbf{a}(\bar{\theta}_{dl,i})
	}{
		\left\|
		\mathbf{a}(\bar{\theta}_{dl,i})
		\right\|
	},
\end{equation}
$\bar{\theta}_{dl,i}$ denotes the predicted direction from DL AP $i$
toward the target hotspot, and $P_{\rm AP}$ is the maximum transmit
power available at each AP.

The received signal at UL AP $m\in\mathcal{M}_{\rm UL}$ during the
$t$-th sensing symbol is
\begin{equation}
	\mathbf{y}_m^{s}[t]
	=
	\sum_{i\in\mathcal{M}_{\rm DL}}
	\mathbf{H}_{m,i}^{\rm AA}\mathbf{x}_i^{s}[t]
	+
	\delta_T
	\sum_{i\in\mathcal{M}_{\rm DL}}
	\mathbf{H}_{m,i}^{\rm ATA}\mathbf{x}_i^{s}[t]
	+
	\mathbf{n}_m^{s}[t],
	\label{eq:sensing_rx_ap}
\end{equation}
where $\mathbf{H}_{m,i}^{\rm AA}$ is the target-free channel between
DL AP $i$ and UL AP $m$, $\mathbf{H}_{m,i}^{\rm ATA}$ is the
corresponding target-reflected channel defined in
\eqref{eq:ATA_channel}, and
\begin{equation}
	\mathbf{n}_m^{s}[t]
	\sim
	\mathcal{CN}
	\left(
	\mathbf{0},
	\sigma_{\rm ul}^{2}\mathbf{I}_N
	\right)
\end{equation}
denotes the receiver noise at UL AP $m$.

\subsection{Vectorized Cooperative Sensing Model}

To obtain a compact representation of the received sensing signals,
we vectorize the inter-AP channel matrices as
\begin{equation}
	\boldsymbol{\eta}_{m,i}^{\rm AA}
	\triangleq
	\operatorname{vec}
	\left\{
	\mathbf{H}_{m,i}^{\rm AA}
	\right\},
\end{equation}
and
\begin{equation}
	\boldsymbol{\eta}_{m,i}^{\rm ATA}
	\triangleq
	\operatorname{vec}
	\left\{
	\mathbf{H}_{m,i}^{\rm ATA}
	\right\}.
\end{equation}

The target-free vectorized channel is modeled as
\begin{equation}
	\boldsymbol{\eta}_{m,i}^{\rm AA}
	\sim
	\mathcal{CN}
	\left(
	\mathbf{0},
	\mathbf{R}_{m,i}^{\rm AA}
	\right),
\end{equation}
where $\mathbf{R}_{m,i}^{\rm AA}$ denotes its covariance matrix.

Similarly, owing to uncertainty in the predicted target location and
target reflection characteristics, the target-reflected channel is
modeled as
\begin{equation}
	\boldsymbol{\eta}_{m,i}^{\rm ATA}
	\sim
	\mathcal{CN}
	\left(
	\overline{\boldsymbol{\eta}}_{m,i}^{\rm ATA},
	\mathbf{R}_{m,i}^{\rm ATA}
	\right),
	\label{eq:ATA_stat_model}
\end{equation}
where
\begin{equation}
	\overline{\boldsymbol{\eta}}_{m,i}^{\rm ATA}
	\triangleq
	\operatorname{vec}
	\left\{
	\overline{\mathbf{H}}_{m,i}^{\rm ATA}
	\right\}
\end{equation}
is the vectorized mean target-reflected channel, and
$\mathbf{R}_{m,i}^{\rm ATA}$ characterizes the uncertainty associated
with the target-reflected channel.

The mean target-reflected channel follows the physical channel model
introduced in Section~\ref{sec:channel_model}, i.e.,
\begin{equation}
	\overline{\mathbf{H}}_{m,i}^{\rm ATA}
	=
	\widehat{\vartheta}_{m,i}^{\rm RCS}
	\overline{\lambda}_{m,i}^{\rm ATA}
	\mathbf{a}(\bar{\theta}_{ul,m})
	\mathbf{a}^{H}(\bar{\theta}_{dl,i}),
	\label{eq:mean_ATA_channel}
\end{equation}
where $\widehat{\vartheta}_{m,i}^{\rm RCS}$ denotes the estimated
bistatic RCS and the remaining quantities follow the definitions given
in the target-reflected channel model.

The received signals collected by all $M_{\rm UL}$ UL APs during
symbol $t$ are forwarded to the CPU and stacked as
\begin{equation}
	\mathbf{y}^{s}[t]
	=
	\begin{bmatrix}
		(\mathbf{y}_{1}^{s}[t])^{T} &
		\cdots &
		(\mathbf{y}_{M_{\rm UL}}^{s}[t])^{T}
	\end{bmatrix}^{T}.
\end{equation}

Using the vectorization identity
$\operatorname{vec}(\mathbf{A}\mathbf{B}\mathbf{C})
=(\mathbf{C}^{T}\otimes\mathbf{A})\operatorname{vec}(\mathbf{B})$,
the aggregate received sensing signal can be expressed as
\begin{equation}
	\boxed{
		\mathbf{y}^{s}[t]
		=
		\mathbf{X}^{s}[t]\boldsymbol{\eta}^{\rm AA}
		+
		\delta_T
		\mathbf{X}^{s}[t]\boldsymbol{\eta}^{\rm ATA}
		+
		\mathbf{n}^{s}[t]
	}
	\label{eq:aggregate_sensing}
\end{equation}
where
\begin{equation}
	\mathbf{X}^{s}[t]
	=
	\mathbf{I}_{M_{\rm UL}}
	\otimes
	\left(
	\begin{bmatrix}
		(\mathbf{x}_{1}^{s}[t])^{T} &
		\cdots &
		(\mathbf{x}_{M_{\rm DL}}^{s}[t])^{T}
	\end{bmatrix}
	\otimes
	\mathbf{I}_{N}
	\right).
\end{equation}

The aggregate target-free channel vector is defined as
\begin{equation}
	\boldsymbol{\eta}^{\rm AA}
	=
	\begin{bmatrix}
		(\boldsymbol{\eta}_{1,1}^{\rm AA})^{T} &
		\cdots &
		(\boldsymbol{\eta}_{1,M_{\rm DL}}^{\rm AA})^{T} &
		\cdots &
		(\boldsymbol{\eta}_{M_{\rm UL},M_{\rm DL}}^{\rm AA})^{T}
	\end{bmatrix}^{T},
\end{equation}
with
\begin{equation}
	\boldsymbol{\eta}^{\rm AA}
	\sim
	\mathcal{CN}
	\left(
	\mathbf{0},
	\mathbf{R}_{\eta}^{\rm AA}
	\right).
\end{equation}

Likewise,
\begin{equation}
	\boldsymbol{\eta}^{\rm ATA}
	=
	\begin{bmatrix}
		(\boldsymbol{\eta}_{1,1}^{\rm ATA})^{T} &
		\cdots &
		(\boldsymbol{\eta}_{1,M_{\rm DL}}^{\rm ATA})^{T} &
		\cdots &
		(\boldsymbol{\eta}_{M_{\rm UL},M_{\rm DL}}^{\rm ATA})^{T}
	\end{bmatrix}^{T},
\end{equation}
and
\begin{equation}
	\boldsymbol{\eta}^{\rm ATA}
	\sim
	\mathcal{CN}
	\left(
	\overline{\boldsymbol{\eta}}^{\rm ATA},
	\mathbf{R}_{\eta}^{\rm ATA}
	\right).
\end{equation}

Assuming statistically independent AP-to-AP links, the corresponding
aggregate covariance matrices can be represented in block-diagonal
form as
\begin{align}
	\mathbf{R}_{\eta}^{\rm AA}
	&=
	\operatorname{blkdiag}
	\left\{
	\mathbf{R}_{1,1}^{\rm AA},
	\ldots,
	\mathbf{R}_{M_{\rm UL},M_{\rm DL}}^{\rm AA}
	\right\},
	\\
	\mathbf{R}_{\eta}^{\rm ATA}
	&=
	\operatorname{blkdiag}
	\left\{
	\mathbf{R}_{1,1}^{\rm ATA},
	\ldots,
	\mathbf{R}_{M_{\rm UL},M_{\rm DL}}^{\rm ATA}
	\right\}.
\end{align}

The mean vector of the aggregate target-reflected channel is
\begin{equation}
	\overline{\boldsymbol{\eta}}^{\rm ATA}
	=
	\begin{bmatrix}
		(\overline{\boldsymbol{\eta}}_{1,1}^{\rm ATA})^{T} &
		\cdots &
		(\overline{\boldsymbol{\eta}}_{M_{\rm UL},M_{\rm DL}}^{\rm ATA})^{T}
	\end{bmatrix}^{T}.
\end{equation}

\subsection{MAPRT-Based Target Detection}

To determine the target-presence indicator, the CPU employs a
maximum-a-posteriori ratio test (MAPRT). The received sensing signals
over the $\tau_s$ sensing symbols are stacked as
\begin{equation}
	\mathbf{y}_{\tau}
	=
	\begin{bmatrix}
		(\mathbf{y}^{s}[1])^{T} &
		\cdots &
		(\mathbf{y}^{s}[\tau_s])^{T}
	\end{bmatrix}^{T}.
\end{equation}

Similarly, define
\begin{align}
	\mathbf{g}_{\tau}^{\rm AA}
	&=
	\begin{bmatrix}
		(\mathbf{X}^{s}[1]\boldsymbol{\eta}^{\rm AA})^{T} &
		\cdots &
		(\mathbf{X}^{s}[\tau_s]\boldsymbol{\eta}^{\rm AA})^{T}
	\end{bmatrix}^{T},
	\\
	\mathbf{g}_{\tau}^{\rm ATA}
	&=
	\begin{bmatrix}
		(\mathbf{X}^{s}[1]\boldsymbol{\eta}^{\rm ATA})^{T} &
		\cdots &
		(\mathbf{X}^{s}[\tau_s]\boldsymbol{\eta}^{\rm ATA})^{T}
	\end{bmatrix}^{T},
\end{align}
and let $\mathbf{n}_{\tau}$ denote the correspondingly stacked receiver
noise vector.

The binary target-detection problem is then formulated as
\begin{equation}
	\begin{aligned}
		\mathcal{H}_0:\quad
		&\mathbf{y}_{\tau}
		=
		\mathbf{g}_{\tau}^{\rm AA}
		+
		\mathbf{n}_{\tau},
		\\
		\mathcal{H}_1:\quad
		&\mathbf{y}_{\tau}
		=
		\mathbf{g}_{\tau}^{\rm AA}
		+
		\mathbf{g}_{\tau}^{\rm ATA}
		+
		\mathbf{n}_{\tau},
	\end{aligned}
	\label{eq:MAPRT_hypotheses}
\end{equation}
where $\mathcal{H}_0$ represents target absence and $\mathcal{H}_1$
represents target presence.

The joint MAP estimation problem can therefore be expressed as
\begin{equation}
	\left(
	\widehat{\boldsymbol{\eta}}^{\rm AA},
	\widehat{\boldsymbol{\eta}}^{\rm ATA},
	\widehat{\mathcal H}
	\right)
	=
	\underset{
		\boldsymbol{\eta}^{\rm AA},
		\boldsymbol{\eta}^{\rm ATA},
		\mathcal H
	}{
		\arg\max
	}
	\;
	p
	\left(
	\boldsymbol{\eta}^{\rm AA},
	\boldsymbol{\eta}^{\rm ATA},
	\mathcal H
	\mid
	\mathbf{y}_{\tau}
	\right).
	\label{eq:MAP_joint}
\end{equation}

Assuming that the target-free and target-reflected channel components
are statistically independent, the MAPRT decision rule is written as
\begin{equation}
	\boxed{
		\Lambda
		=
		\frac{\Lambda_{\rm num}}{\Lambda_{\rm den}}
		\underset{\mathcal H_0}{\overset{\mathcal H_1}{\gtrless}}
		\Lambda_{\rm det,Th}
	}
	\label{eq:MAPRT}
\end{equation}
where $\Lambda_{\rm det,Th}$ denotes the detection threshold.

The numerator and denominator of the likelihood ratio are respectively
given by
\begin{align}
	\Lambda_{\rm num}
	&=
	\max_{\boldsymbol{\eta}^{\rm AA},
		\boldsymbol{\eta}^{\rm ATA}}
	p
	\left(
	\mathbf{y}_{\tau}
	\mid
	\boldsymbol{\eta}^{\rm AA},
	\boldsymbol{\eta}^{\rm ATA},
	\mathcal H_1
	\right)
	p
	\left(
	\boldsymbol{\eta}^{\rm AA}\mid\mathcal H_1
	\right)
	p
	\left(
	\boldsymbol{\eta}^{\rm ATA}\mid\mathcal H_1
	\right),
	\\
	\Lambda_{\rm den}
	&=
	\max_{\boldsymbol{\eta}^{\rm AA}}
	p
	\left(
	\mathbf{y}_{\tau}
	\mid
	\boldsymbol{\eta}^{\rm AA},
	\mathcal H_0
	\right)
	p
	\left(
	\boldsymbol{\eta}^{\rm AA}\mid\mathcal H_0
	\right).
\end{align}

The threshold $\Lambda_{\rm det,Th}$ can be selected according to a
prescribed false-alarm probability or other desired target-detection
criterion. Finally, the estimated target-presence indicator is obtained
as
\begin{equation}
	\boxed{
		\widehat{\delta}_T
		=
		\begin{cases}
			1, & \Lambda > \Lambda_{\rm det,Th},\\[1mm]
			0, & \Lambda \leq \Lambda_{\rm det,Th}.
		\end{cases}
	}
	\label{eq:target_indicator_est}
\end{equation}
\section{Channel Estimation in the Downlink Pilot Training Stage}
\label{sec:dl_pilot_training}

During the downlink pilot training stage, all antennas of the
$M_{\rm DL}$ DL APs transmit mutually orthogonal pilot sequences, while
the $M_{\rm UL}$ UL APs jointly receive these pilots and estimate the
inter-AP channels with the aid of the target-presence estimate
$\widehat{\delta}_T$ obtained in the preceding target detection stage.

Let
\[
\tau_{\rm dp}=M_{\rm DL}N
\]
denote the number of mutually orthogonal downlink pilot sequences.
The pilot sequence assigned to the $n$-th antenna of DL AP
$i\in\mathcal{M}_{\rm DL}$ is denoted by
$\boldsymbol{\phi}_{i(n)}^{\rm dp}\in\mathbb{C}^{\tau_{\rm dp}}$,
where the pilot sequences satisfy
\begin{equation}
	\left(\boldsymbol{\phi}_{i(n)}^{\rm dp}\right)^H
	\boldsymbol{\phi}_{j(q)}^{\rm dp}
	=
	\begin{cases}
		\tau_{\rm dp}, & i=j,\; n=q,\\
		0, & \text{otherwise}.
	\end{cases}
	\label{eq:dl_pilot_orthogonality}
\end{equation}

For DL AP $i$ and UL AP $m$, the effective inter-AP channel during the
current coherence interval is defined as
\begin{equation}
	\boxed{
		\mathbf H_{m,i}^{\rm J}
		\triangleq
		\mathbf H_{m,i}^{\rm AA}
		+
		\delta_T\mathbf H_{m,i}^{\rm ATA}
	}
	\label{eq:joint_interAP_channel}
\end{equation}
where $\mathbf H_{m,i}^{\rm AA}\in\mathbb{C}^{N\times N}$ denotes the
target-free inter-AP channel and
$\mathbf H_{m,i}^{\rm ATA}\in\mathbb{C}^{N\times N}$ denotes the
target-reflected channel introduced in the preceding section.

Equivalently,
\begin{equation}
	\mathbf H_{m,i}^{\rm J}
	=
	\begin{bmatrix}
		\mathbf h_{m,i(1)}^{\rm J} &
		\cdots &
		\mathbf h_{m,i(N)}^{\rm J}
	\end{bmatrix}
	\in\mathbb{C}^{N\times N},
	\label{eq:joint_interAP_columns}
\end{equation}
where
\begin{equation}
	\mathbf h_{m,i(n)}^{\rm J}
	=
	\mathbf h_{m,i(n)}^{\rm AA}
	+
	\delta_T\mathbf h_{m,i(n)}^{\rm ATA}
	\in\mathbb{C}^{N\times1}
	\label{eq:joint_channel_column}
\end{equation}
denotes the effective channel from the $n$-th antenna of DL AP $i$ to
UL AP $m$.

\subsection{Received Downlink Pilot Signal}

The pilot signal received at UL AP $m\in\mathcal{M}_{\rm UL}$ is
expressed as
\begin{equation}
	\mathbf Y_m^{\rm dp}
	=
	\sqrt{\tau_{\rm dp}\rho_{\rm dp}}
	\sum_{i\in\mathcal{M}_{\rm DL}}
	\sum_{n=1}^{N}
	\mathbf h_{m,i(n)}^{\rm J}
	\left(
	\boldsymbol{\phi}_{i(n)}^{\rm dp}
	\right)^H
	+
	\mathbf N_m^{\rm dp},
	\label{eq:received_dl_pilot}
\end{equation}
where $\rho_{\rm dp}$ denotes the downlink pilot transmit power and
$\mathbf N_m^{\rm dp}$ is the receiver-noise matrix whose entries are
independent complex Gaussian random variables with variance
$\sigma_{\rm ul}^{2}$.

To estimate the effective channel
$\mathbf h_{m,i(n)}^{\rm J}$, UL AP $m$ correlates the received pilot
matrix with the pilot assigned to the $n$-th antenna of DL AP $i$.
Accordingly, the sufficient statistic is
\begin{equation}
	\mathbf y_{m,i(n)}^{\rm dp}
	\triangleq
	\mathbf Y_m^{\rm dp}
	\boldsymbol{\phi}_{i(n)}^{\rm dp}.
	\label{eq:dl_pilot_projection}
\end{equation}

Owing to the orthogonality of the downlink pilot sequences,
$\mathbf y_{m,i(n)}^{\rm dp}$ provides the observation required for
estimating the corresponding inter-AP channel.

\subsection{MMSE Estimation of the Inter-AP Channel}

Following the MMSE estimation principle, the estimate of
$\mathbf h_{m,i(n)}^{\rm J}$ can be expressed in its general form as
\begin{align}
	\widehat{\mathbf h}_{m,i(n)}^{\rm J}
	={}&
	\mathbb{E}
	\left\{
	\mathbf h_{m,i(n)}^{\rm J}
	\right\}
	\nonumber\\
	&+
	\operatorname{Cov}
	\left\{
	\mathbf h_{m,i(n)}^{\rm J},
	\mathbf y_{m,i(n)}^{\rm dp}
	\right\}
	\operatorname{Cov}
	\left\{
	\mathbf y_{m,i(n)}^{\rm dp},
	\mathbf y_{m,i(n)}^{\rm dp}
	\right\}^{-1}
	\nonumber\\
	&\quad\times
	\left(
	\mathbf y_{m,i(n)}^{\rm dp}
	-
	\mathbb{E}
	\left\{
	\mathbf y_{m,i(n)}^{\rm dp}
	\right\}
	\right).
	\label{eq:general_mmse_interAP}
\end{align}

From \eqref{eq:joint_channel_column}, the conditional mean of the
effective inter-AP channel is determined by the target-reflected
component. Let
\begin{equation}
	\overline{\mathbf h}_{m,i(n)}^{\rm ATA}
	\triangleq
	\mathbb{E}
	\left\{
	\mathbf h_{m,i(n)}^{\rm ATA}
	\right\}.
\end{equation}
Then,
\begin{equation}
	\mathbb{E}
	\left\{
	\mathbf h_{m,i(n)}^{\rm J}
	\right\}
	=
	\delta_T
	\overline{\mathbf h}_{m,i(n)}^{\rm ATA}.
	\label{eq:joint_channel_mean}
\end{equation}

For compactness, define the MMSE filtering matrix
$\mathbf S_{m,i(n)}^{\rm J}$ according to the second-order statistics
of the effective inter-AP channel and the corresponding pilot
observation. The MMSE estimate can then be written as
\begin{align}
	\widehat{\mathbf h}_{m,i(n)}^{\rm J}
	={}&
	\delta_T
	\left(
	\mathbf I_N
	-
	\sqrt{\tau_{\rm dp}\rho_{\rm dp}}\,
	\mathbf S_{m,i(n)}^{\rm J}
	\right)
	\overline{\mathbf h}_{m,i(n)}^{\rm ATA}
	\nonumber\\
	&+
	\mathbf S_{m,i(n)}^{\rm J}
	\mathbf y_{m,i(n)}^{\rm dp}.
	\label{eq:mmse_interAP_true_delta}
\end{align}

\subsection{Target-Detection-Assisted Channel Estimate}

The true target indicator $\delta_T$ is unknown to the CPU.
Consequently, the CPU employs the estimated target indicator
$\widehat{\delta}_T$ obtained from the MAPRT detector in
Section~\ref{sec:target_detection}. The practical estimate of the
inter-AP channel is therefore defined as
\begin{align}
	\check{\mathbf h}_{m,i(n)}^{\rm J}
	={}&
	\widehat{\delta}_T
	\left(
	\mathbf I_N
	-
	\sqrt{\tau_{\rm dp}\rho_{\rm dp}}\,
	\mathbf S_{m,i(n)}^{\rm J}
	\right)
	\overline{\mathbf h}_{m,i(n)}^{\rm ATA}
	\nonumber\\
	&+
	\mathbf S_{m,i(n)}^{\rm J}
	\mathbf y_{m,i(n)}^{\rm dp}.
	\label{eq:practical_interAP_estimate}
\end{align}

The actual effective inter-AP channel follows
\begin{equation}
	\mathbf h_{m,i(n)}^{\rm J}
	\sim
	\mathcal{CN}
	\left(
	\delta_T
	\overline{\mathbf h}_{m,i(n)}^{\rm ATA},
	\mathbf R_{m,i(n)}^{\rm J}
	\right),
	\label{eq:true_interAP_distribution}
\end{equation}
where $\mathbf R_{m,i(n)}^{\rm J}$ denotes the covariance matrix of the
effective inter-AP channel.

Correspondingly, the target-detection-assisted channel estimate is
modeled as
\begin{equation}
	\check{\mathbf h}_{m,i(n)}^{\rm J}
	\sim
	\mathcal{CN}
	\left(
	\widehat{\delta}_T
	\overline{\mathbf h}_{m,i(n)}^{\rm ATA},
	\mathbf\Phi_{m,i(n)}^{\rm J}
	\right),
	\label{eq:estimated_interAP_distribution}
\end{equation}
where $\mathbf\Phi_{m,i(n)}^{\rm J}$ denotes the covariance matrix of
the estimated inter-AP channel.
\section{Spectral-Efficiency and Sensing-SINR Analysis in the
	Data-Transmission and Target-Sensing Stage}
\label{sec:se_sensing_analysis}

\subsection{Downlink Spectral-Efficiency Analysis}
\label{subsec:dl_se}

During the data-transmission and target-sensing stage, the
$M_{\rm DL}$ DL APs jointly transmit communication signals to the
$K_{\rm DL}$ DL UEs. When the target is detected, the DL APs also
transmit dedicated sensing waveforms to illuminate the target and
estimate its motion-related parameters. Hence, communication and
sensing are performed simultaneously over the same time-frequency
resources through multi-antenna precoding.

The transmit signal of DL AP $i\in\mathcal{M}_{\rm DL}$ is expressed as
\begin{equation}
	\mathbf{x}_{i}^{\rm DL}
	=
	\sum_{k\in\mathcal{K}_{\rm DL}}
	\mathbf{w}_{i,k}^{\rm com}s_{k}^{\rm DL}
	+
	\widehat{\delta}_{T}\,
	\mathbf{w}_{i}^{\rm sen}s_{i}^{\rm sen},
	\label{eq:DL_tx_signal}
\end{equation}
where
$\mathbf{w}_{i,k}^{\rm com}\in\mathbb{C}^{N\times1}$ denotes the
communication beamforming vector employed by DL AP $i$ for DL UE $k$,
and $s_k^{\rm DL}$ is the corresponding unit-power information symbol,
i.e.,
\begin{equation}
	\mathbb{E}
	\left\{
	|s_k^{\rm DL}|^2
	\right\}=1.
\end{equation}

Moreover, $s_i^{\rm sen}$ denotes the dedicated sensing symbol
transmitted by DL AP $i$, which is statistically independent of the
communication data symbols and satisfies
\begin{equation}
	\mathbb{E}
	\left\{
	|s_i^{\rm sen}|^2
	\right\}
	=1,
\end{equation}
and
\begin{equation}
	\mathbb{E}
	\left\{
	s_i^{\rm sen}
	(s_{i'}^{\rm sen})^{*}
	\right\}
	=0,
	\qquad i\neq i'.
\end{equation}
The corresponding sensing beamforming vector is denoted by
$\mathbf{w}_{i}^{\rm sen}\in\mathbb{C}^{N\times1}$.

The binary quantity $\widehat{\delta}_{T}$ represents the target
indicator obtained from the MAPRT-based target-detection stage.
Therefore, the sensing transmission is activated only when the target
is declared present.

Using the channel decomposition introduced in the uplink pilot
training stage,
\begin{equation}
	\mathbf h_{i,l}^{\rm UA}
	=
	\widehat{\mathbf h}_{i,l}^{\rm UA}
	+
	\widetilde{\mathbf h}_{i,l}^{\rm UA},
	\label{eq:DL_channel_decomposition_again}
\end{equation}
the received signal at DL UE $l\in\mathcal{K}_{\rm DL}$ can be
decomposed as
\begin{align}
	y_l^{\rm DL}
	={}&
	\underbrace{
		\sum_{i\in\mathcal{M}_{\rm DL}}
		\left(
		\widehat{\mathbf h}_{i,l}^{\rm UA}
		\right)^H
		\mathbf w_{i,l}^{\rm com}
		s_l^{\rm DL}
	}_{D_l^{\rm DL}}
	\nonumber\\
	&+
	\underbrace{
		\sum_{\substack{k\in\mathcal{K}_{\rm DL}\\k\neq l}}
		\sum_{i\in\mathcal{M}_{\rm DL}}
		\left(
		\widehat{\mathbf h}_{i,l}^{\rm UA}
		\right)^H
		\mathbf w_{i,k}^{\rm com}
		s_k^{\rm DL}
	}_{I_l^{\rm MU,DL}}
	\nonumber\\
	&+
	\underbrace{
		\sum_{k\in\mathcal{K}_{\rm DL}}
		\sum_{i\in\mathcal{M}_{\rm DL}}
		\left(
		\widetilde{\mathbf h}_{i,l}^{\rm UA}
		\right)^H
		\mathbf w_{i,k}^{\rm com}
		s_k^{\rm DL}
	}_{I_l^{\rm CE,DL}}
	\nonumber\\
	&+
	\underbrace{
		\widehat{\delta}_{T}
		\sum_{i\in\mathcal{M}_{\rm DL}}
		\left(
		\mathbf h_{i,l}^{\rm UA}
		\right)^H
		\mathbf w_i^{\rm sen}
		s_i^{\rm sen}
	}_{I_l^{\rm sen,DL}}
	\nonumber\\
	&+
	\underbrace{
		\sum_{u\in\mathcal{K}_{\rm UL}}
		\sqrt{p_u}\,
		h_{u,l}^{\rm UU}
		s_u^{\rm UL}
	}_{I_l^{\rm CLI,DL}}
	+
	n_l^{\rm DL},
	\label{eq:DL_received_signal}
\end{align}
where $p_u$ denotes the transmit power of UL UE $u$,
$s_u^{\rm UL}$ is its unit-power UL data symbol, and
\begin{equation}
	h_{u,l}^{\rm UU}\in\mathbb C
\end{equation}
denotes the UL-UE-to-DL-UE cross-link channel. The effect of path loss,
shadowing, and small-scale fading is directly incorporated into
$h_{u,l}^{\rm UU}$, so that no separate channel-gain coefficient is
required.

The additive receiver noise is modeled as
\begin{equation}
	n_l^{\rm DL}
	\sim
	\mathcal{CN}
	\left(
	0,\sigma_{\rm DL}^{2}
	\right).
\end{equation}

In \eqref{eq:DL_received_signal}, $D_l^{\rm DL}$ denotes the desired
DL communication signal, whereas
$I_l^{\rm MU,DL}$,
$I_l^{\rm CE,DL}$,
$I_l^{\rm sen,DL}$, and
$I_l^{\rm CLI,DL}$ respectively represent the multi-user DL
interference, channel-estimation-error-induced interference,
sensing-induced interference, and UL-to-DL cross-link interference.

Accordingly, the instantaneous DL SINR of UE $l$ is written as
\begin{equation}
	\Gamma_l^{\rm DL}
	=
	\frac{
		\left|
		D_l^{\rm DL}
		\right|^2
	}{
		\left|
		I_l^{\rm MU,DL}
		\right|^2
		+
		\left|
		I_l^{\rm CE,DL}
		\right|^2
		+
		\left|
		I_l^{\rm sen,DL}
		\right|^2
		+
		\left|
		I_l^{\rm CLI,DL}
		\right|^2
		+
		\sigma_{\rm DL}^{2}
	}.
	\label{eq:DL_SINR_general}
\end{equation}
\subsection{Uplink Spectral-Efficiency Analysis}
\label{subsec:ul_se}

During uplink transmission, the $K_{\rm UL}$ UL UEs simultaneously
transmit their data symbols to the $M_{\rm UL}$ UL APs. Meanwhile, the
DL APs continue transmitting the joint communication and sensing
signals introduced in \eqref{eq:DL_tx_signal}. Consequently, the
received signal at UL AP $m\in\mathcal{M}_{\rm UL}$ is expressed as
\begin{equation}
	\mathbf y_m^{\rm UL}
	=
	\sum_{u\in\mathcal{K}_{\rm UL}}
	\sqrt{p_u}\,
	\mathbf h_{m,u}^{\rm UA}s_u^{\rm UL}
	+
	\sum_{i\in\mathcal{M}_{\rm DL}}
	\mathbf H_{m,i}^{\rm J}\mathbf x_i^{\rm DL}
	+
	\mathbf n_m^{\rm UL},
	\label{eq:UL_received_before_CLI}
\end{equation}
where $p_u$ denotes the transmit power of UL UE $u$,
$s_u^{\rm UL}$ is its unit-power information symbol satisfying
\begin{equation}
	\mathbb E\{|s_u^{\rm UL}|^2\}=1,
\end{equation}
and $\mathbf h_{m,u}^{\rm UA}\in\mathbb C^{N\times1}$ denotes the
channel between UL UE $u$ and UL AP $m$. The receiver noise is modeled
as
\begin{equation}
	\mathbf n_m^{\rm UL}
	\sim
	\mathcal{CN}
	\left(
	\mathbf 0,
	\sigma_{\rm UL}^{2}\mathbf I_N
	\right).
\end{equation}

The second term in \eqref{eq:UL_received_before_CLI} represents the
DL-to-UL inter-AP cross-link interference (CLI). Owing to the relatively
high DL transmit power and the dense deployment of APs, this
interference can severely degrade the UL reception performance if it is
not properly suppressed.

In the proposed interference-management framework, the CPU has access
to both the DL transmit signals and the target-detection-assisted
inter-AP channel estimates obtained during the downlink pilot training
stage. Specifically, the practical estimate of the effective inter-AP
channel is written as
\begin{equation}
	\check{\mathbf H}_{m,i}^{\rm J}
	=
	\begin{bmatrix}
		\check{\mathbf h}_{m,i(1)}^{\rm J} &
		\cdots &
		\check{\mathbf h}_{m,i(N)}^{\rm J}
	\end{bmatrix},
	\label{eq:estimated_J_matrix}
\end{equation}
where $\check{\mathbf h}_{m,i(n)}^{\rm J}$ was defined in
\eqref{eq:practical_interAP_estimate}.

The actual effective inter-AP channel can therefore be decomposed as
\begin{equation}
	\mathbf H_{m,i}^{\rm J}
	=
	\check{\mathbf H}_{m,i}^{\rm J}
	+
	\widetilde{\mathbf H}_{m,i}^{\rm J},
	\label{eq:J_channel_error_decomposition}
\end{equation}
where
\begin{equation}
	\widetilde{\mathbf H}_{m,i}^{\rm J}
	\triangleq
	\mathbf H_{m,i}^{\rm J}
	-
	\check{\mathbf H}_{m,i}^{\rm J}
\end{equation}
denotes the residual inter-AP channel estimation error.

Since $\check{\mathbf H}_{m,i}^{\rm J}$ and
$\mathbf x_i^{\rm DL}$ are available at the CPU, the reconstructed
DL-to-UL interference
\begin{equation}
	\sum_{i\in\mathcal M_{\rm DL}}
	\check{\mathbf H}_{m,i}^{\rm J}\mathbf x_i^{\rm DL}
\end{equation}
can be subtracted from the received signal. Hence, after inter-AP CLI
cancellation, the residual signal at UL AP $m$ becomes
\begin{equation}
	\widetilde{\mathbf y}_m^{\rm UL}
	=
	\sum_{u\in\mathcal K_{\rm UL}}
	\sqrt{p_u}\,
	\mathbf h_{m,u}^{\rm UA}s_u^{\rm UL}
	+
	\sum_{i\in\mathcal M_{\rm DL}}
	\widetilde{\mathbf H}_{m,i}^{\rm J}
	\mathbf x_i^{\rm DL}
	+
	\mathbf n_m^{\rm UL}.
	\label{eq:UL_received_after_CLI}
\end{equation}

Using the uplink channel decomposition
\begin{equation}
	\mathbf h_{m,u}^{\rm UA}
	=
	\widehat{\mathbf h}_{m,u}^{\rm UA}
	+
	\widetilde{\mathbf h}_{m,u}^{\rm UA},
	\label{eq:UL_channel_decomposition}
\end{equation}
the CPU detects the data transmitted by UL UE $j$ by combining the
signals received at all UL APs. The resulting decision statistic is
\begin{equation}
	r_j^{\rm UL}
	=
	\sum_{m\in\mathcal M_{\rm UL}}
	\mathbf v_{m,j}^{H}
	\widetilde{\mathbf y}_m^{\rm UL},
	\label{eq:UL_combined_signal}
\end{equation}
where $\mathbf v_{m,j}\in\mathbb C^{N\times1}$ denotes the receive
combining vector associated with UL UE $j$ at UL AP $m$.

Following the considered MRC receiver, the combining vector is selected
as
\begin{equation}
	\mathbf v_{m,j}
	=
	\widehat{\mathbf h}_{m,j}^{\rm UA}.
	\label{eq:MRC_vector}
\end{equation}

Substituting \eqref{eq:UL_received_after_CLI},
\eqref{eq:UL_channel_decomposition}, and \eqref{eq:DL_tx_signal} into
\eqref{eq:UL_combined_signal}, the detected signal can be decomposed as
\begin{align}
	r_j^{\rm UL}
	={}&
	\underbrace{
		\sum_{m\in\mathcal M_{\rm UL}}
		\sqrt{p_j}\,
		\mathbf v_{m,j}^{H}
		\widehat{\mathbf h}_{m,j}^{\rm UA}
		s_j^{\rm UL}
	}_{D_j^{\rm UL}}
	\nonumber\\
	&+
	\underbrace{
		\sum_{m\in\mathcal M_{\rm UL}}
		\mathbf v_{m,j}^{H}
		\left(
		\sum_{\substack{u\in\mathcal K_{\rm UL}\\u\neq j}}
		\sqrt{p_u}\,
		\widehat{\mathbf h}_{m,u}^{\rm UA}
		s_u^{\rm UL}
		\right)
	}_{I_j^{\rm MU,UL}}
	\nonumber\\
	&+
	\underbrace{
		\sum_{m\in\mathcal M_{\rm UL}}
		\mathbf v_{m,j}^{H}
		\left(
		\sum_{u\in\mathcal K_{\rm UL}}
		\sqrt{p_u}\,
		\widetilde{\mathbf h}_{m,u}^{\rm UA}
		s_u^{\rm UL}
		\right)
	}_{I_j^{\rm CE,UL}}
	\nonumber\\
	&+
	\underbrace{
		\sum_{m\in\mathcal M_{\rm UL}}
		\mathbf v_{m,j}^{H}
		\left[
		\sum_{i\in\mathcal M_{\rm DL}}
		\widetilde{\mathbf H}_{m,i}^{\rm J}
		\left(
		\sum_{k\in\mathcal K_{\rm DL}}
		\mathbf w_{i,k}^{\rm com}s_k^{\rm DL}
		\right)
		\right]
	}_{I_j^{\rm CLI,UL}}
	\nonumber\\
	&+
	\underbrace{
		\widehat{\delta}_T
		\sum_{m\in\mathcal M_{\rm UL}}
		\mathbf v_{m,j}^{H}
		\left(
		\sum_{i\in\mathcal M_{\rm DL}}
		\widetilde{\mathbf H}_{m,i}^{\rm J}
		\mathbf w_i^{\rm sen}s_i^{\rm sen}
		\right)
	}_{I_j^{\rm sen,UL}}
	\nonumber\\
	&+
	\underbrace{
		\sum_{m\in\mathcal M_{\rm UL}}
		\mathbf v_{m,j}^{H}
		\mathbf n_m^{\rm UL}
	}_{I_j^{\rm N,UL}}.
	\label{eq:UL_signal_decomposition}
\end{align}

In \eqref{eq:UL_signal_decomposition}, $D_j^{\rm UL}$ denotes the
desired signal component of UL UE $j$.
The term $I_j^{\rm MU,UL}$ represents the multi-user UL interference
generated by the remaining UL UEs, whereas $I_j^{\rm CE,UL}$ accounts
for the interference caused by UL channel-estimation errors.
Furthermore, $I_j^{\rm CLI,UL}$ represents the residual DL-to-UL
inter-AP CLI generated by the DL communication signals after
interference reconstruction and cancellation, while
$I_j^{\rm sen,UL}$ represents the residual interference associated
with the sensing waveform. Finally, $I_j^{\rm N,UL}$ denotes the
post-combining receiver-noise component.

Accordingly, the instantaneous UL SINR of UL UE $j$ is expressed as
\begin{equation}
	\boxed{
		\Gamma_j^{\rm UL}
		=
		\frac{
			|D_j^{\rm UL}|^2
		}{
			|I_j^{\rm MU,UL}|^2
			+
			|I_j^{\rm CE,UL}|^2
			+
			|I_j^{\rm CLI,UL}|^2
			+
			|I_j^{\rm sen,UL}|^2
			+
			|I_j^{\rm N,UL}|^2
		}.
	}
	\label{eq:UL_SINR}
\end{equation}
\subsection{Sensing SINR Analysis}
\label{subsec:sensing_sinr}

Based on the received signal in \eqref{eq:UL_received_before_CLI}, the
signal observed at UL AP $m$ can be decomposed, from the sensing
perspective, as
\begin{equation}
	\mathbf y_m^{\rm UL}
	=
	\mathbf{DS}_m^{\rm sen}
	+
	\mathbf I_m^{\rm U,sen}
	+
	\mathbf I_m^{\rm D,sen}
	+
	\mathbf n_m^{\rm UL},
	\label{eq:sensing_received_decomposition}
\end{equation}
where the desired target-reflected sensing signal is given by
\begin{equation}
	\mathbf{DS}_m^{\rm sen}
	=
	\widehat{\delta}_T
	\sum_{i\in\mathcal M_{\rm DL}}
	\mathbf H_{m,i}^{\rm ATA}
	\mathbf w_i^{\rm sen}s_i^{\rm sen},
	\label{eq:desired_sensing_signal}
\end{equation}
while
\begin{equation}
	\mathbf I_m^{\rm U,sen}
	=
	\sum_{u\in\mathcal K_{\rm UL}}
	\sqrt{p_u}\,
	\mathbf h_{m,u}^{\rm UA}s_u^{\rm UL}
	\label{eq:UL_comm_sensing_interference}
\end{equation}
and
\begin{equation}
	\mathbf I_m^{\rm D,sen}
	=
	\sum_{i\in\mathcal M_{\rm DL}}
	\mathbf H_{m,i}^{\rm J}
	\left(
	\sum_{k\in\mathcal K_{\rm DL}}
	\mathbf w_{i,k}^{\rm com}s_k^{\rm DL}
	\right)
	\label{eq:DL_comm_sensing_interference}
\end{equation}
represent the interference components caused by the UL and DL
communication signals, respectively. These interference components
can significantly degrade the sensing performance. Nevertheless, the
proposed interference-management mechanism exploits the available CSI
and decoded communication symbols to partially reconstruct and suppress
them.

For the UL communication interference, the channel
$\mathbf h_{m,u}^{\rm UA}$ has already been estimated during the uplink
pilot training stage. Moreover, after UL data detection, the CPU can
obtain an estimate $\widehat{s}_u^{\rm UL}$ of the transmitted symbol
$s_u^{\rm UL}$. Therefore, the UL communication interference can be
reconstructed and subtracted from the received sensing signal. The
corresponding residual UL interference is expressed as
\begin{equation}
	\widetilde{\mathbf I}_m^{\rm U,sen}
	=
	\sum_{u\in\mathcal K_{\rm UL}}
	\sqrt{p_u}\,
	\mathbf h_{m,u}^{\rm UA}
	\widetilde{s}_u^{\rm UL},
	\label{eq:residual_UL_sensing_interference}
\end{equation}
where
\begin{equation}
	\widetilde{s}_u^{\rm UL}
	\triangleq
	s_u^{\rm UL}-\widehat{s}_u^{\rm UL}
	\label{eq:UL_symbol_error}
\end{equation}
denotes the UL symbol-detection error. Following the considered error
model, it is characterized as
\begin{equation}
	\widetilde{s}_u^{\rm UL}
	\sim
	\mathcal{CN}
	\left(0,\sigma_{\widetilde{s}_u^{\rm UL}}^2\right),
\end{equation}
where $\sigma_{\widetilde{s}_u^{\rm UL}}^2$ decreases as the
corresponding UL SINR increases.

Similarly, the CPU can exploit the estimated effective inter-AP
channel $\check{\mathbf H}_{m,i}^{\rm J}$ together with the known DL
communication signals to reconstruct the DL communication interference.
Accordingly, after interference reconstruction and cancellation, the
residual DL communication interference is given by
\begin{equation}
	\widetilde{\mathbf I}_m^{\rm D,sen}
	=
	\sum_{i\in\mathcal M_{\rm DL}}
	\widetilde{\mathbf H}_{m,i}^{\rm J}
	\left(
	\sum_{k\in\mathcal K_{\rm DL}}
	\mathbf w_{i,k}^{\rm com}s_k^{\rm DL}
	\right),
	\label{eq:residual_DL_sensing_interference}
\end{equation}
where
\begin{equation}
	\widetilde{\mathbf H}_{m,i}^{\rm J}
	\triangleq
	\mathbf H_{m,i}^{\rm J}
	-
	\check{\mathbf H}_{m,i}^{\rm J}
	\label{eq:sensing_interAP_error}
\end{equation}
denotes the residual inter-AP channel error.

After subtracting the reconstructed UL and DL communication
interference components, the residual sensing observation at UL AP $m$
can therefore be expressed as
\begin{equation}
	\widetilde{\mathbf y}_m^{\rm sen}
	=
	\mathbf{DS}_m^{\rm sen}
	+
	\widetilde{\mathbf I}_m^{\rm U,sen}
	+
	\widetilde{\mathbf I}_m^{\rm D,sen}
	+
	\mathbf n_m^{\rm UL}.
	\label{eq:residual_sensing_received}
\end{equation}

The sensing observation is subsequently processed using a receive
filter $\mathbf u_m\in\mathbb C^{N\times1}$ at UL AP $m$. The
corresponding scalar radar observation is
\begin{equation}
	r_m^{\rm sen}
	=
	\mathbf u_m^{H}
	\widetilde{\mathbf y}_m^{\rm sen}.
	\label{eq:radar_filtered_signal}
\end{equation}
Hence, for given communication and sensing precoders, the output
sensing SINR at UL AP $m$ is expressed as
\begin{equation}
	\gamma_m^{\rm sen}
	=
	\frac{
		\mathbf u_m^{H}
		\mathbf{DS}_m^{\rm sen}
		(\mathbf{DS}_m^{\rm sen})^{H}
		\mathbf u_m
	}{
		\mathbf u_m^{H}
		\mathbf Q_m
		\mathbf u_m
	},
	\label{eq:sensing_sinr}
\end{equation}
where
\begin{equation}
	\mathbf Q_m
	\triangleq
	\left(
	\widetilde{\mathbf I}_m^{\rm U,sen}
	+
	\widetilde{\mathbf I}_m^{\rm D,sen}
	+
	\mathbf n_m^{\rm UL}
	\right)
	\left(
	\widetilde{\mathbf I}_m^{\rm U,sen}
	+
	\widetilde{\mathbf I}_m^{\rm D,sen}
	+
	\mathbf n_m^{\rm UL}
	\right)^{H}
	\label{eq:sensing_interference_covariance}
\end{equation}
denotes the interference-plus-noise covariance matrix.

The receive filter $\mathbf u_m$ has a direct impact on the achievable
sensing SINR. Therefore, for fixed communication and sensing
beamformers, it can be designed by maximizing
\eqref{eq:sensing_sinr}, i.e.,
\begin{equation}
	\mathbf u_m^{\star}
	=
	\arg\max_{\mathbf u_m\neq\mathbf 0}
	\frac{
		\mathbf u_m^{H}
		\mathbf{DS}_m^{\rm sen}
		(\mathbf{DS}_m^{\rm sen})^{H}
		\mathbf u_m
	}{
		\mathbf u_m^{H}\mathbf Q_m\mathbf u_m
	}.
	\label{eq:sensing_filter_optimization}
\end{equation}
The above problem has the form of a generalized Rayleigh quotient.
Consequently, an optimal receive filter, up to an arbitrary nonzero
scaling factor, is given by
\begin{equation}
	\boxed{
		\mathbf u_m^{\star}
		=
		\mathbf Q_m^{-1}
		\mathbf{DS}_m^{\rm sen}.
	}
	\label{eq:optimal_sensing_filter}
\end{equation}

Substituting \eqref{eq:optimal_sensing_filter} into
\eqref{eq:sensing_sinr}, the maximum achievable sensing SINR at UL AP
$m$ becomes
\begin{equation}
	\boxed{
		\gamma_m^{\rm sen,\star}
		=
		(\mathbf{DS}_m^{\rm sen})^{H}
		\mathbf Q_m^{-1}
		\mathbf{DS}_m^{\rm sen}.
	}
	\label{eq:maximum_sensing_sinr}
\end{equation}
Accordingly, $\gamma_m^{\rm sen,\star}$ is adopted as the sensing
performance metric at UL AP $m$.

%=========================================================
\section{Priority-Based and Joint Beamforming Strategies}
\label{sec:beamforming_strategies}
%=========================================================

To characterize the sensing--communication tradeoff of the
considered NAFD cell-free ISAC system, we introduce two
priority-based beamforming strategies in addition to the proposed
joint design. The first strategy prioritizes communication and
designs the sensing beam such that its interference to the
communication links is suppressed. The second strategy prioritizes
sensing and subsequently optimizes the communication beamformers
without degrading the predetermined sensing beam.

For these baseline schemes, the maximum transmit power of DL AP
$i$ is divided between communication and sensing according to
\begin{equation}
	P_i^{\mathrm{c}}
	=
	\rho P_i^{\max},
	\qquad
	P_i^{\mathrm{s}}
	=
	(1-\rho)P_i^{\max},
	\label{eq:priority_power_split}
\end{equation}
where $0\leq\rho\leq1$ denotes the fraction of the DL AP power
allocated to communication.

Let
\begin{equation}
	\widehat{\mathbf H}_{\mathrm d}
	=
	\begin{bmatrix}
		\widehat{\mathbf h}_1 &
		\cdots &
		\widehat{\mathbf h}_{K_{\mathrm d}}
	\end{bmatrix}
	\in
	\mathbb C^{N_{\mathrm{tx}}\times K_{\mathrm d}}
	\label{eq:Hdl_priority}
\end{equation}
denote the aggregated estimated DL channel matrix.

Furthermore, define
\begin{equation}
	\mathbf c_j
	=
	\widetilde{\mathbf H}_{\mathrm{du}}^{H}
	\mathbf v_j,
	\qquad
	j\in\mathcal K_{\mathrm u},
	\label{eq:effective_du_channel_priority}
\end{equation}
as the effective residual DL-to-UL interference channel observed
after the UL receive combiner $\mathbf v_j$.

The combined communication-interference channel matrix is defined as
\begin{equation}
	\mathbf C
	=
	\begin{bmatrix}
		\widehat{\mathbf H}_{\mathrm d} &
		\mathbf c_1 &
		\cdots &
		\mathbf c_{K_{\mathrm u}}
	\end{bmatrix}.
	\label{eq:Cpriority}
\end{equation}

Finally, let $\mathbf a_{\mathrm s}$ denote the dominant sensing
direction. For the general sensing matrix $\mathbf A$, it is selected
as
\begin{equation}
	\mathbf a_{\mathrm s}
	=
	\mathbf u_{\max}(\mathbf A),
	\label{eq:dominant_sensing_direction}
\end{equation}
where $\mathbf u_{\max}(\mathbf A)$ denotes the normalized eigenvector
associated with the largest eigenvalue of $\mathbf A$.

%=========================================================
\subsection{Communication-Prioritized Sensing Beamforming}
\label{subsec:comm_priority}
%=========================================================

In the communication-prioritized strategy, the communication
beamforming vectors are designed first. The sensing beam is then
constructed such that it does not introduce additional interference
to either the DL UEs or the UL communication receivers.

The communication beamformers may, for example, be obtained using
regularized zero forcing (RZF). Define
\begin{equation}
	\widetilde{\mathbf w}_k^{\mathrm{RZF}}
	=
	\left(
	\widehat{\mathbf H}_{\mathrm d}
	\widehat{\mathbf H}_{\mathrm d}^{H}
	+
	\lambda_{\mathrm R}\mathbf I
	\right)^{-1}
	\widehat{\mathbf h}_k,
	\label{eq:RZF_priority}
\end{equation}
where $\lambda_{\mathrm R}>0$ is the regularization factor.

The beam directions are normalized according to
\begin{equation}
	\overline{\mathbf w}_k^{\mathrm{RZF}}
	=
	\frac{
		\widetilde{\mathbf w}_k^{\mathrm{RZF}}
	}{
		\left\|
		\widetilde{\mathbf w}_k^{\mathrm{RZF}}
		\right\|_2
	}.
\end{equation}

To prevent the sensing signal from degrading communication, define
the orthogonal projector onto the nullspace of the effective
communication channels as
\begin{equation}
	\mathbf P_{\perp}
	=
	\mathbf I
	-
	\mathbf C
	\left(
	\mathbf C^{H}\mathbf C
	\right)^{\dagger}
	\mathbf C^{H}.
	\label{eq:communication_nullspace}
\end{equation}

The unnormalized communication-prioritized sensing beam is
\begin{equation}
	\widetilde{\mathbf w}_{\mathrm s}^{\mathrm{CP}}
	=
	\mathbf P_{\perp}
	\mathbf a_{\mathrm s}.
	\label{eq:CP_sensing_raw}
\end{equation}

Consequently,
\begin{equation}
	\widehat{\mathbf h}_k^{H}
	\widetilde{\mathbf w}_{\mathrm s}^{\mathrm{CP}}
	=
	0,
	\qquad
	\forall k\in\mathcal K_{\mathrm d},
	\label{eq:CP_DL_null}
\end{equation}
and
\begin{equation}
	\mathbf v_j^{H}
	\widetilde{\mathbf H}_{\mathrm{du}}
	\widetilde{\mathbf w}_{\mathrm s}^{\mathrm{CP}}
	=
	0,
	\qquad
	\forall j\in\mathcal K_{\mathrm u}.
	\label{eq:CP_UL_null}
\end{equation}

Hence, the sensing beam produces neither direct sensing interference
at the DL UEs nor residual DL-to-UL sensing interference after UL
combining.

To satisfy the sensing-power budget of every DL AP, define
\begin{equation}
	\xi_{\mathrm{CP}}
	=
	\min_{i\in\mathcal M_{\mathrm d}}
	\sqrt{
		\frac{
			P_i^{\mathrm s}
		}{
			\left(
			\widetilde{\mathbf w}_{\mathrm s}^{\mathrm{CP}}
			\right)^{H}
			\mathbf D_i
			\widetilde{\mathbf w}_{\mathrm s}^{\mathrm{CP}}
		}
	}.
	\label{eq:CP_scaling}
\end{equation}

The final sensing beam is therefore
\begin{equation}
	\mathbf w_{\mathrm s}^{\mathrm{CP}}
	=
	\xi_{\mathrm{CP}}
	\widetilde{\mathbf w}_{\mathrm s}^{\mathrm{CP}}.
	\label{eq:CP_sensing_beam}
\end{equation}

This baseline gives communication the highest priority and uses only
the remaining spatial degrees of freedom for target illumination.

%=========================================================
\subsection{Sensing-Prioritized Communication Beamforming}
\label{subsec:sensing_priority}
%=========================================================

In the sensing-prioritized strategy, the sensing beam is designed
first without considering the communication channels. The
communication beamformers are subsequently optimized while the
sensing beam is kept fixed.

The unconstrained sensing-optimal direction is
$\mathbf a_{\mathrm s}$. To satisfy all per-AP sensing-power
constraints, define
\begin{equation}
	\xi_{\mathrm{SP}}
	=
	\min_{i\in\mathcal M_{\mathrm d}}
	\sqrt{
		\frac{
			P_i^{\mathrm s}
		}{
			\mathbf a_{\mathrm s}^{H}
			\mathbf D_i
			\mathbf a_{\mathrm s}
		}
	}.
	\label{eq:SP_scaling}
\end{equation}

The sensing-prioritized beam is therefore
\begin{equation}
	\mathbf w_{\mathrm s}^{\mathrm{SP}}
	=
	\xi_{\mathrm{SP}}
	\mathbf a_{\mathrm s}.
	\label{eq:SP_sensing_beam}
\end{equation}

For fixed $\mathbf w_{\mathrm s}^{\mathrm{SP}}$ and fixed UL powers
$\mathbf p$, the communication beamformers are obtained by solving
the max--min DL SINR problem
\begin{subequations}
	\label{prob:SP_comm}
	\begin{align}
		\mathcal P_{\mathrm{SP}}:\quad
		\max_{\{\mathbf w_k\},\,\gamma}
		\quad&
		\gamma
		\label{prob:SP_obj}
		\\
		\text{s.t.}\quad&
		\Gamma_k^{\mathrm d}
		\geq
		\gamma,
		\qquad
		\forall k\in\mathcal K_{\mathrm d},
		\label{prob:SP_DL}
		\\
		&
		\Gamma_j^{\mathrm u}
		\geq
		\gamma_j^{\mathrm u},
		\qquad
		\forall j\in\mathcal K_{\mathrm u},
		\label{prob:SP_UL}
		\\
		&
		\sum_{k\in\mathcal K_{\mathrm d}}
		\mathbf w_k^{H}
		\mathbf D_i
		\mathbf w_k
		\leq
		P_i^{\mathrm c},
		\qquad
		\forall i\in\mathcal M_{\mathrm d}.
		\label{prob:SP_power}
	\end{align}
\end{subequations}

For a trial value $\gamma$, the DL constraint can be converted to an
SOC constraint. Define
\begin{equation}
	\zeta_k^{\mathrm{ud}}
	=
	\sum_{j\in\mathcal K_{\mathrm u}}
	p_j\eta_{j,k}
	+
	\sigma_{\mathrm d,k}^{2}.
	\label{eq:SP_ud_noise}
\end{equation}

After choosing the phase of $\mathbf w_k$ such that
\begin{equation}
	\operatorname{Im}
	\left\{
	\widehat{\mathbf h}_k^{H}\mathbf w_k
	\right\}
	=0,
\end{equation}
the DL SINR requirement is equivalently written as
\begin{align}
	&
	\sqrt{1+\frac{1}{\gamma}}\,
	\operatorname{Re}
	\left\{
	\widehat{\mathbf h}_k^{H}\mathbf w_k
	\right\}
	\nonumber\\
	&\qquad\geq
	\left\|
	\begin{bmatrix}
		\widehat{\mathbf h}_k^{H}\mathbf w_1\\
		\vdots\\
		\widehat{\mathbf h}_k^{H}\mathbf w_{K_{\mathrm d}}\\
		\widehat{\mathbf h}_k^{H}
		\mathbf w_{\mathrm s}^{\mathrm{SP}}\\
		\sqrt{\zeta_k^{\mathrm{ud}}}
	\end{bmatrix}
	\right\|_2,
	\qquad
	\forall k.
	\label{eq:SP_SOC_DL}
\end{align}

The UL QoS constraint can similarly be expressed as
\begin{align}
	&
	\left\|
	\begin{bmatrix}
		\mathbf c_j^{H}\mathbf w_1\\
		\vdots\\
		\mathbf c_j^{H}\mathbf w_{K_{\mathrm d}}\\
		\mathbf c_j^{H}\mathbf w_{\mathrm s}^{\mathrm{SP}}
	\end{bmatrix}
	\right\|_2^2
	\nonumber\\
	&\qquad\leq
	\frac{
		p_j a_{j,j}
	}{
		\gamma_j^{\mathrm u}
	}
	-
	\sum_{\substack{\ell\in\mathcal K_{\mathrm u}\\\ell\neq j}}
	p_{\ell}a_{j,\ell}
	-
	\sigma_{\mathrm u}^{2}
	\|\mathbf v_j\|_2^2.
	\label{eq:SP_UL_constraint}
\end{align}

Therefore, the maximum common communication SINR can be obtained by
bisection over $\gamma$ while solving the corresponding convex
feasibility problem.

For centralized optimization, use an aggregated combiner
$\mathbf u$ and define the effective target matrix
$\mathbf H_{\mathrm t}$. The sensing SINR is represented as
\begin{equation}
	\Gamma_{\mathrm s}
	=
	\frac{
		\widehat{\delta}_{\mathrm T}
		|\mathbf u^{H}\mathbf H_{\mathrm t}\mathbf w_{\mathrm s}|^{2}
	}{
		I_{\mathrm s}
	},
	\label{eq:global_sensing_sinr}
\end{equation}
with
\begin{align}
	I_{\mathrm s}
	={}&
	\sum_{k\in\mathcal K_{\mathrm d}}
	|\mathbf u^{H}
	\widetilde{\mathbf H}_{\mathrm t}\mathbf w_k|^{2}
	+
	\sum_{j\in\mathcal K_{\mathrm u}}p_j\delta_j
	+
	\sigma_{\mathrm s}^{2}.
	\label{eq:global_sensing_interference}
\end{align}
The matrix $\widetilde{\mathbf H}_{\mathrm t}$ captures the
residual DL communication leakage caused by inter-AP CSI
errors, while $\delta_j$ captures the residual UL-to-sensing
interference including decoding and channel-estimation errors.
%=========================================================
\section{Vector-Form Joint Optimization}
%=========================================================

The proposed design maximizes sensing performance while
jointly optimizing the DL communication beamformers, the
sensing beamformer, and the UL powers. The receive combiners
are treated as fixed during each transmit-variable update and
are refreshed using the estimated CSI.

The vector-form problem is
\begin{subequations}
\label{prob:vector}
\begin{align}
\mathcal P_{\mathrm V}:\quad
\underset{\{\mathbf w_k\},\mathbf w_{\mathrm s},\mathbf p}
{\operatorname{maximize}}
\quad&
\Gamma_{\mathrm s}
\label{prob:vector_obj}
\\
\operatorname{subject~to}\quad
&
\Gamma_k^{\mathrm d}
\geq\gamma_k^{\mathrm d},
\quad \forall k\in\mathcal K_{\mathrm d},
\label{prob:vector_dl}
\\
&
\Gamma_j^{\mathrm u}
\geq\gamma_j^{\mathrm u},
\quad \forall j\in\mathcal K_{\mathrm u},
\label{prob:vector_ul}
\\
&
\sum_{k\in\mathcal K_{\mathrm d}}
\|\mathbf w_{i,k}\|_2^2
+
\widehat{\delta}_{\mathrm T}
\|\mathbf w_{i,\mathrm s}\|_2^2
\leq P_i^{\max},
\quad \forall i\in\mathcal M_{\mathrm d},
\label{prob:vector_ap_power}
\\
&
0\leq p_j\leq P_j^{\max},
\quad \forall j\in\mathcal K_{\mathrm u}.
\label{prob:vector_ul_power}
\end{align}
\end{subequations}
When $\widehat{\delta}_{\mathrm T}=0$, the sensing beam is
switched off and the system operates in communication-only
mode. When $\widehat{\delta}_{\mathrm T}=1$, the full joint
communication-and-sensing design is performed.

%=========================================================
\section{Matrix SDR Reformulation}
%=========================================================

Define
\begin{equation}
\mathbf W_k=\mathbf w_k\mathbf w_k^{H},
\qquad
\mathbf W_{\mathrm s}=\mathbf w_{\mathrm s}\mathbf w_{\mathrm s}^{H}.
\end{equation}
Then
\begin{align}
\mathbf W_k&\succeq\mathbf 0,
&\operatorname{rank}(\mathbf W_k)&=1,
\\
\mathbf W_{\mathrm s}&\succeq\mathbf 0,
&\operatorname{rank}(\mathbf W_{\mathrm s})&=1.
\end{align}
Define
\begin{align}
\widehat{\mathbf H}_k
&=
\widehat{\mathbf h}_{k}\widehat{\mathbf h}_{k}^{H},
\\
\mathbf B_j
&=
\widetilde{\mathbf H}_{\mathrm{du}}^{H}
\mathbf v_j\mathbf v_j^{H}
\widetilde{\mathbf H}_{\mathrm{du}},
\\
\mathbf A
&=
\widehat{\delta}_{\mathrm T}
\mathbf H_{\mathrm t}^{H}
\mathbf u\mathbf u^{H}
\mathbf H_{\mathrm t},
\\
\mathbf A_{\mathrm I}
&=
\widetilde{\mathbf H}_{\mathrm t}^{H}
\mathbf u\mathbf u^{H}
\widetilde{\mathbf H}_{\mathrm t},
\\
\eta_{j,k}
&=|g_{j,k}^{\mathrm{ud}}|^{2},
\\
 a_{j,q}
&=|\mathbf v_j^{H}\widehat{\mathbf g}_{q}|^{2},
\\
 c_{j,q}
&=\mathbf v_j^{H}\mathbf C_q^{\mathrm u}\mathbf v_j.
\end{align}
The DL SINR becomes
\begin{equation}
\Gamma_k^{\mathrm d}
=
\frac{
\operatorname{Tr}(\widehat{\mathbf H}_k\mathbf W_k)
}{
\overline I_k^{\mathrm d}
},
\end{equation}
where
\begin{align}
\overline I_k^{\mathrm d}
={}&
\sum_{\ell\neq k}
\operatorname{Tr}(\widehat{\mathbf H}_k\mathbf W_{\ell})
+
\widehat{\delta}_{\mathrm T}
\operatorname{Tr}(\widehat{\mathbf H}_k\mathbf W_{\mathrm s})
\nonumber\\
&+
\sum_{\ell\in\mathcal K_{\mathrm d}}
\operatorname{Tr}(\mathbf C_k^{\mathrm d}\mathbf W_{\ell})
+
\widehat{\delta}_{\mathrm T}
\operatorname{Tr}(\mathbf C_k^{\mathrm d}\mathbf W_{\mathrm s})
\nonumber\\
&+
\sum_{j\in\mathcal K_{\mathrm u}}p_j\eta_{j,k}
+
\sigma_{\mathrm d,k}^{2}.
\label{eq:matrix_dl_denominator}
\end{align}
The UL SINR becomes
\begin{equation}
\Gamma_j^{\mathrm u}
=
\frac{p_j a_{j,j}}{\overline I_j^{\mathrm u}},
\end{equation}
where
\begin{align}
\overline I_j^{\mathrm u}
={}&
\sum_{q\neq j}p_q a_{j,q}
+
\sum_{q\in\mathcal K_{\mathrm u}}p_qc_{j,q}
\nonumber\\
&+
\sum_{k\in\mathcal K_{\mathrm d}}
\operatorname{Tr}(\mathbf B_j\mathbf W_k)
+
\widehat{\delta}_{\mathrm T}
\operatorname{Tr}(\mathbf B_j\mathbf W_{\mathrm s})
+
\sigma_{\mathrm u}^{2}\|\mathbf v_j\|_2^{2}.
\label{eq:matrix_ul_denominator}
\end{align}
The sensing SINR is
\begin{equation}
\Gamma_{\mathrm s}
=
\frac{
\operatorname{Tr}(\mathbf A\mathbf W_{\mathrm s})
}{
\displaystyle
\sum_{k\in\mathcal K_{\mathrm d}}
\operatorname{Tr}(\mathbf A_{\mathrm I}\mathbf W_k)
+
\sum_{j\in\mathcal K_{\mathrm u}}p_j\delta_j
+
\sigma_{\mathrm s}^{2}
}.
\label{eq:matrix_sensing_sinr}
\end{equation}
Let $\mathbf D_i$ be the diagonal selection matrix that extracts
the antennas of DL AP $i$. The matrix problem is
\begin{subequations}
\label{prob:matrix}
\begin{align}
\mathcal P_{\mathrm M}:\quad
\underset{\{\mathbf W_k\},\mathbf W_{\mathrm s},\mathbf p}
{\operatorname{maximize}}
\quad&
\Gamma_{\mathrm s}
\\
\operatorname{subject~to}\quad
&
\Gamma_k^{\mathrm d}\geq\gamma_k^{\mathrm d},
\quad\forall k,
\\
&
\Gamma_j^{\mathrm u}\geq\gamma_j^{\mathrm u},
\quad\forall j,
\\
&
\sum_{k\in\mathcal K_{\mathrm d}}
\operatorname{Tr}(\mathbf D_i\mathbf W_k)
+
\widehat{\delta}_{\mathrm T}
\operatorname{Tr}(\mathbf D_i\mathbf W_{\mathrm s})
\leq P_i^{\max},
\quad\forall i,
\\
&
0\leq p_j\leq P_j^{\max},
\quad\forall j,
\\
&
\mathbf W_k\succeq\mathbf 0,
\quad\operatorname{rank}(\mathbf W_k)=1,
\quad\forall k,
\\
&
\mathbf W_{\mathrm s}\succeq\mathbf 0,
\quad\operatorname{rank}(\mathbf W_{\mathrm s})=1.
\end{align}
\end{subequations}
The semidefinite relaxation drops the rank-one constraints.

%=========================================================
\section{SDR-Based Alternating Optimization}
%=========================================================

\subsection{Beamforming Update}

For fixed $\mathbf p^{(t)}$, introduce a trial sensing SINR
$\tau$. The sensing constraint is
\begin{align}
\operatorname{Tr}(\mathbf A\mathbf W_{\mathrm s})
\geq{}&
\tau
\sum_{k\in\mathcal K_{\mathrm d}}
\operatorname{Tr}(\mathbf A_{\mathrm I}\mathbf W_k)
\nonumber\\
&+
\tau\left(
\sum_{j\in\mathcal K_{\mathrm u}}p_j^{(t)}\delta_j
+
\sigma_{\mathrm s}^{2}
\right).
\label{eq:sdr_sensing_constraint}
\end{align}
The DL QoS constraints become linear in the covariance
matrices:
\begin{align}
\operatorname{Tr}(\widehat{\mathbf H}_k\mathbf W_k)
\geq{}&
\gamma_k^{\mathrm d}
\sum_{\ell\neq k}
\operatorname{Tr}(\widehat{\mathbf H}_k\mathbf W_{\ell})
\nonumber\\
&+
\gamma_k^{\mathrm d}
\widehat{\delta}_{\mathrm T}
\operatorname{Tr}(\widehat{\mathbf H}_k\mathbf W_{\mathrm s})
\nonumber\\
&+
\gamma_k^{\mathrm d}
\sum_{\ell}
\operatorname{Tr}(\mathbf C_k^{\mathrm d}\mathbf W_{\ell})
\nonumber\\
&+
\gamma_k^{\mathrm d}
\widehat{\delta}_{\mathrm T}
\operatorname{Tr}(\mathbf C_k^{\mathrm d}\mathbf W_{\mathrm s})
\nonumber\\
&+
\gamma_k^{\mathrm d}
\left(
\sum_jp_j^{(t)}\eta_{j,k}
+
\sigma_{\mathrm d,k}^{2}
\right).
\label{eq:sdr_dl_constraint}
\end{align}
The UL QoS constraints become
\begin{align}
&
\sum_{k\in\mathcal K_{\mathrm d}}
\operatorname{Tr}(\mathbf B_j\mathbf W_k)
+
\widehat{\delta}_{\mathrm T}
\operatorname{Tr}(\mathbf B_j\mathbf W_{\mathrm s})
\nonumber\\
&\leq
\frac{p_j^{(t)}a_{j,j}}{\gamma_j^{\mathrm u}}
-
\sum_{q\neq j}p_q^{(t)}a_{j,q}
-
\sum_qp_q^{(t)}c_{j,q}
-
\sigma_{\mathrm u}^{2}\|\mathbf v_j\|_2^{2}.
\label{eq:sdr_ul_constraint}
\end{align}
For fixed $\tau$, these constraints, the per-AP power
constraints, and the PSD constraints form an SDP feasibility
problem. Bisection on $\tau$ gives the beamforming update.

\subsection{UL Power Update}

For fixed covariance matrices, maximizing the sensing SINR is
equivalent to minimizing the residual UL-to-sensing
interference:
\begin{subequations}
\label{prob:power_update}
\begin{align}
\underset{\mathbf p}{\operatorname{minimize}}
\quad&
\sum_{j\in\mathcal K_{\mathrm u}}\delta_jp_j
\\
\operatorname{subject~to}\quad
&
\sum_{j\in\mathcal K_{\mathrm u}}p_j\eta_{j,k}
\leq\zeta_k,
\quad\forall k,
\\
&
p_ja_{j,j}
-
\gamma_j^{\mathrm u}
\sum_{q\neq j}p_qa_{j,q}
-
\gamma_j^{\mathrm u}
\sum_qp_qc_{j,q}
\geq
\gamma_j^{\mathrm u}\xi_j,
\quad\forall j,
\\
&
0\leq p_j\leq P_j^{\max},
\quad\forall j,
\end{align}
\end{subequations}
where
\begin{align}
\zeta_k
={}&
\frac{\operatorname{Tr}(\widehat{\mathbf H}_k\mathbf W_k)}
{\gamma_k^{\mathrm d}}
-
\sum_{\ell\neq k}
\operatorname{Tr}(\widehat{\mathbf H}_k\mathbf W_{\ell})
\nonumber\\
&-
\widehat{\delta}_{\mathrm T}
\operatorname{Tr}(\widehat{\mathbf H}_k\mathbf W_{\mathrm s})
-
\sum_{\ell}
\operatorname{Tr}(\mathbf C_k^{\mathrm d}\mathbf W_{\ell})
\nonumber\\
&-
\widehat{\delta}_{\mathrm T}
\operatorname{Tr}(\mathbf C_k^{\mathrm d}\mathbf W_{\mathrm s})
-
\sigma_{\mathrm d,k}^{2},
\\
\xi_j
={}&
\sum_{k\in\mathcal K_{\mathrm d}}
\operatorname{Tr}(\mathbf B_j\mathbf W_k)
+
\widehat{\delta}_{\mathrm T}
\operatorname{Tr}(\mathbf B_j\mathbf W_{\mathrm s})
+
\sigma_{\mathrm u}^{2}\|\mathbf v_j\|_2^{2}.
\end{align}
The power update is a linear program.

%=========================================================
\begin{algorithm}[t]
\caption{Target-Aware SDR Alternating Optimization}
\label{alg:target_aware_sdr}
\begin{algorithmic}[1]
\Require Pilot observations, sensing observations, QoS
thresholds, power budgets, tolerances $\epsilon_{\mathrm{AO}}$
and $\epsilon_{\mathrm{bis}}$
\Ensure $\{\mathbf w_k^{\star}\}$,
$\mathbf w_{\mathrm s}^{\star}$, and $\mathbf p^{\star}$
\State Estimate UE--AP channels
\State Compute $\widehat{\delta}_{\mathrm T}$ using the MAPRT
\State Estimate inter-AP channels 
\State Form channel-error covariances and residual channel
$\widetilde{\mathbf H}_{\mathrm{du}}$
\If{$\widehat{\delta}_{\mathrm T}=0$}
    \State Set $\mathbf W_{\mathrm s}=\mathbf 0$
\EndIf
\State Initialize a feasible $\mathbf p^{(0)}$ and set $t=0$
\Repeat
    \State Update UL communication combiners
    $\{\mathbf v_j\}$ from the estimated UL CSI
    \State Update sensing combiner $\mathbf u$ using
    \eqref{eq:optimal_sensing_filter}
    \State Fix $\mathbf p=\mathbf p^{(t)}$
    \State Choose bisection bounds $\tau_{\mathrm L}$ and
    $\tau_{\mathrm U}$
    \While{$\tau_{\mathrm U}-\tau_{\mathrm L}>
    \epsilon_{\mathrm{bis}}$}
        \State $\tau\gets(\tau_{\mathrm L}+\tau_{\mathrm U})/2$
        \State Solve the SDP feasibility problem defined by
        \eqref{eq:sdr_sensing_constraint}--
        \eqref{eq:sdr_ul_constraint}
        \If{the SDP is feasible}
            \State $\tau_{\mathrm L}\gets\tau$ and store the
            covariance matrices
        \Else
            \State $\tau_{\mathrm U}\gets\tau$
        \EndIf
    \EndWhile
    \State Set the stored solution to
    $\{\mathbf W_k^{(t+1)}\}$ and
    $\mathbf W_{\mathrm s}^{(t+1)}$
    \State Solve \eqref{prob:power_update} and obtain
    $\mathbf p^{(t+1)}$
    \State Evaluate $\Gamma_{\mathrm s}^{(t+1)}$
    \State $t\gets t+1$
\Until{$|\Gamma_{\mathrm s}^{(t)}-
\Gamma_{\mathrm s}^{(t-1)}|\leq\epsilon_{\mathrm{AO}}$}
\State Recover rank-one beamformers by principal-eigenvector
extraction; use Gaussian randomization if required
\State \Return $\{\mathbf w_k^{\star}\}$,
$\mathbf w_{\mathrm s}^{\star}$, and $\mathbf p^{\star}$
\end{algorithmic}
\end{algorithm}
%=========================================================

\section{Remarks on the Proposed Formulation}

The channel-estimation and target-detection stages follow the
four-stage interference-management structure. The direct
interference cancellation in the UL and sensing receivers uses
the estimated inter-AP CSI. The SDR-based optimization is an
extension for jointly optimizing the DL communication beams,
the sensing beam, and the UL powers; it is not the deep-learning
solver used in the reference interference-management paper.
The explicit channel-error covariance terms make the QoS
constraints conservative with respect to imperfect CSI, while
$\widehat{\delta}_{\mathrm T}$ switches the sensing function on or
off according to the detector output.

\section{Numerical Results}
\label{sec:numerical_results}

In this section, the performance of the proposed NAFD cell-free ISAC framework is evaluated through numerical simulations. The proposed alternating-optimization (AO) procedure is implemented in MATLAB, where the beamforming subproblem is solved through semidefinite relaxation (SDR). In addition, the MAPRT detector is employed during the target-detection stage to determine the presence of the sensing target.

\subsection{Simulation Setup}

We consider an NAFD cell-free ISAC network consisting of $M=10$ distributed APs, among which $M_{\mathrm{DL}}=5$ APs operate in the downlink mode and $M_{\mathrm{UL}}=5$ APs operate in the uplink mode. Each AP is equipped with $N_t=N_r=4$ antennas. The network simultaneously serves $K_{\mathrm{DL}}=4$ downlink UEs and $K_{\mathrm{UL}}=4$ uplink UEs. Therefore, the dimensions of the aggregated downlink transmit and uplink receive arrays are
\begin{equation}
	N_{\mathrm{tx}}=M_{\mathrm{DL}}N_t=20,
	\qquad
	N_{\mathrm{rx}}=M_{\mathrm{UL}}N_r=20.
\end{equation}

The APs and UEs are distributed over a square service region with side length $200$~m. A minimum AP--UE separation of $10$~m is considered. The path-loss exponents associated with the AP--UE, UE--UE, and AP--AP links are set to $3.2$, $3.8$, and $2.6$, respectively.

The coherence interval consists of $\tau_c=200$ symbols. The uplink pilot length is $\tau_p=8$, while the default sensing duration is $\tau_s=10$ symbols. For studying the target-detection performance, the sensing duration is varied over
\begin{equation}
	\tau_s\in\{5,8,10,12,16,20,24,32\}.
\end{equation}
The inter-AP pilot length is selected according to
\begin{equation}
	\tau_{\mathrm{DP}}=M_{\mathrm{DL}}N_t=20.
\end{equation}

The maximum transmit power of each DL AP is $P_{\mathrm{AP}}=3$~W, whereas the maximum UL UE transmit power is $P_{\mathrm{UE}}=0.1$~W. The UL pilot power and the inter-AP pilot power are set to $0.1$~W and $3$~W, respectively. The receiver noise power is fixed at $-83$~dBm.

The original minimum communication SINR requirements are set to
\begin{equation}
	\gamma_k^{\mathrm{DL}}
	=
	\gamma_j^{\mathrm{UL}}
	=
	0~\mathrm{dB},
	\qquad \forall k,j.
\end{equation}
In the DL-SINR refinement stage, the optimized sensing solution is retained while a secondary max--min DL-SINR optimization is performed. In the considered realization, this procedure results in a refined common DL SINR target of approximately $2.1$~dB.

For the AO algorithm, the maximum number of iterations is set to $15$, and the convergence tolerance is $\epsilon_{\mathrm{AO}}=10^{-3}$. The bisection tolerance used in the SDR optimization is $10^{-3}$ with a maximum of $20$ bisection iterations.

For target detection, the desired false-alarm probability of the MAPRT detector is selected as
\begin{equation}
	P_{\mathrm{FA}}^{\mathrm{des}}=0.05.
\end{equation}
The detector threshold is calibrated from $300$ target-absent realizations, and the detection probabilities are evaluated using $300$ Monte Carlo trials for each sensing duration. The target-reflection scaling coefficient used in the simulation is $10^{-3}$.

\subsection{Network Realization and Interference Structure}

\begin{figure}[t]
	\centering
	\includegraphics[width=\linewidth]{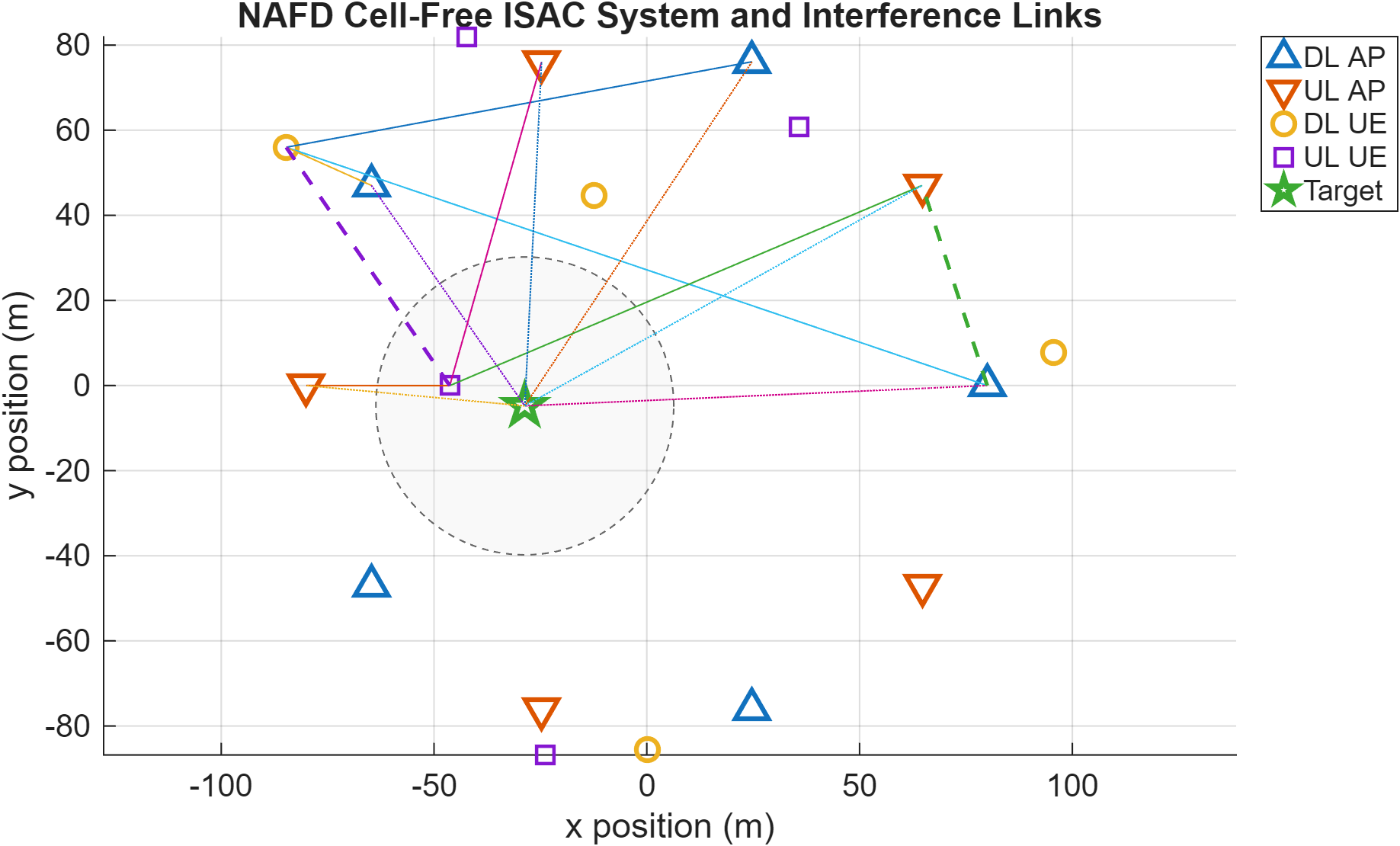}
	\caption{Example realization of the considered NAFD cell-free ISAC network and the associated communication, sensing, and interference links.}
	\label{fig:network}
\end{figure}

Fig.~\ref{fig:network} illustrates one realization of the considered NAFD cell-free ISAC network. The five DL APs and five UL APs are geographically distributed over the service area together with four DL UEs, four UL UEs, and one sensing target. The target is located close to the center of the considered sensing region. The illustrated links represent the simultaneous communication, sensing, and cross-link interference paths included in the system model.

\subsubsection{Impact of the Communication--Sensing Power Allocation}

\begin{figure}[t]
	\centering
	\includegraphics[width=\linewidth]{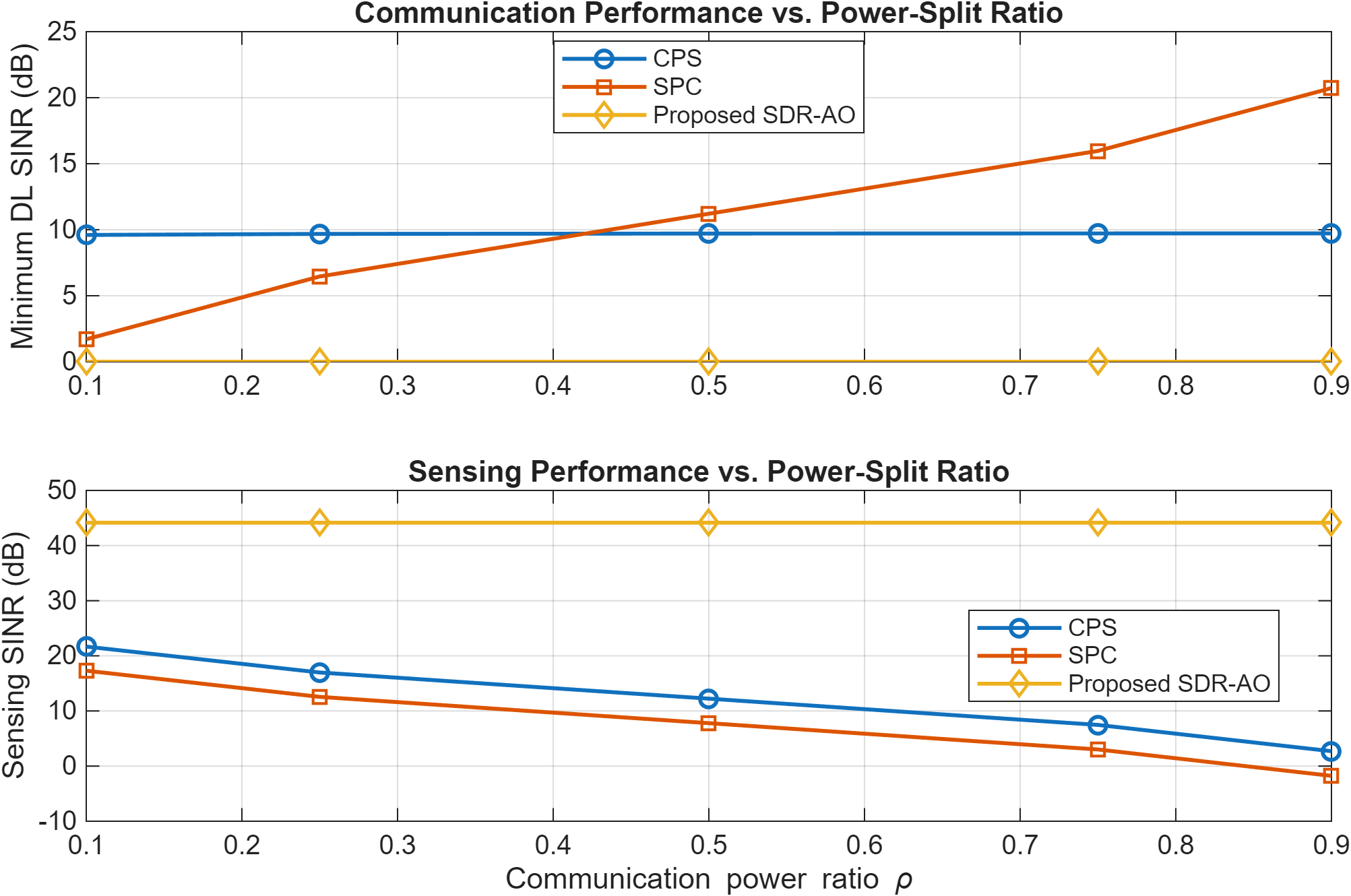}
	\caption{Communication and sensing performance versus the communication
		power ratio $\rho$ for CPS, SPC, and the proposed SDR-AO scheme.}
	\label{fig:power_split_comparison}
	\label{fig:comp}
\end{figure}

Fig.~\ref{fig:comp} investigates the communication--sensing
trade-off as a function of the communication power ratio $\rho$. For the
priority-based schemes, a fraction $\rho$ of the available AP transmit power
is assigned to communication, whereas the remaining fraction $1-\rho$ is
reserved for sensing. The communication-prioritized sensing (CPS) scheme first
constructs the communication beamformers and subsequently designs the sensing
beam in the null space of the communication-related channels. In contrast, the
sensing-prioritized communication (SPC) scheme first determines the sensing
beam and then optimizes the communication beamformers using the remaining
communication power.

As shown in the upper subplot of Fig.~\ref{fig:comp}, the
minimum DL SINR achieved by SPC increases considerably with $\rho$, from
approximately $1.5$~dB at $\rho=0.1$ to more than $20$~dB at $\rho=0.9$.
This behavior is expected because increasing $\rho$ directly increases the
power available for communication beamforming. The CPS scheme exhibits a
different behavior, maintaining a minimum DL SINR of approximately
$9.5$--$9.7$~dB over the considered range of $\rho$. This indicates that,
for the investigated realization, the communication-oriented beam directions
and the corresponding QoS constraints rather than the available transmit
power constitute the dominant limitation of the CPS solution.

The lower subplot demonstrates the complementary sensing behavior. The
sensing SINR of both priority-based approaches decreases as $\rho$ increases,
since a progressively smaller fraction of the AP power remains available for
sensing. In particular, the CPS sensing SINR decreases from approximately
$21$~dB at $\rho=0.1$ to approximately $2$--$3$~dB at $\rho=0.9$, while the
SPC sensing SINR decreases from approximately $17$~dB to below $0$~dB.
Consequently, the priority-based approaches exhibit the expected explicit
trade-off between communication and sensing: allocating additional power to
communication improves, or preserves, the communication performance at the
expense of sensing performance.

In contrast, the proposed SDR-AO solution is not constrained by a predefined
communication--sensing power split. Instead, the communication beamformers,
sensing covariance matrix, and UL transmit powers are jointly optimized under
the AP power and communication QoS constraints. Therefore, its curve is
independent of $\rho$ in this comparison and should be interpreted as a
joint-power-allocation benchmark rather than as another fixed-split solution.
The proposed method achieves a sensing SINR of approximately $44$~dB over the
entire range, substantially exceeding both priority-based approaches. This
large gain shows the benefit of exploiting the available spatial and power
degrees of freedom jointly rather than reserving predetermined portions of the
transmit power for communication and sensing. The corresponding communication
QoS is maintained by the constraints of the proposed optimization problem.
Hence, the results demonstrate that the proposed SDR-AO design can strongly
enhance sensing performance without relying on an externally selected
communication--sensing power-allocation ratio.

\subsubsection{Performance with Individual Communication QoS Constraints}

\begin{figure}[t]
	\centering
	\includegraphics[width=\linewidth]{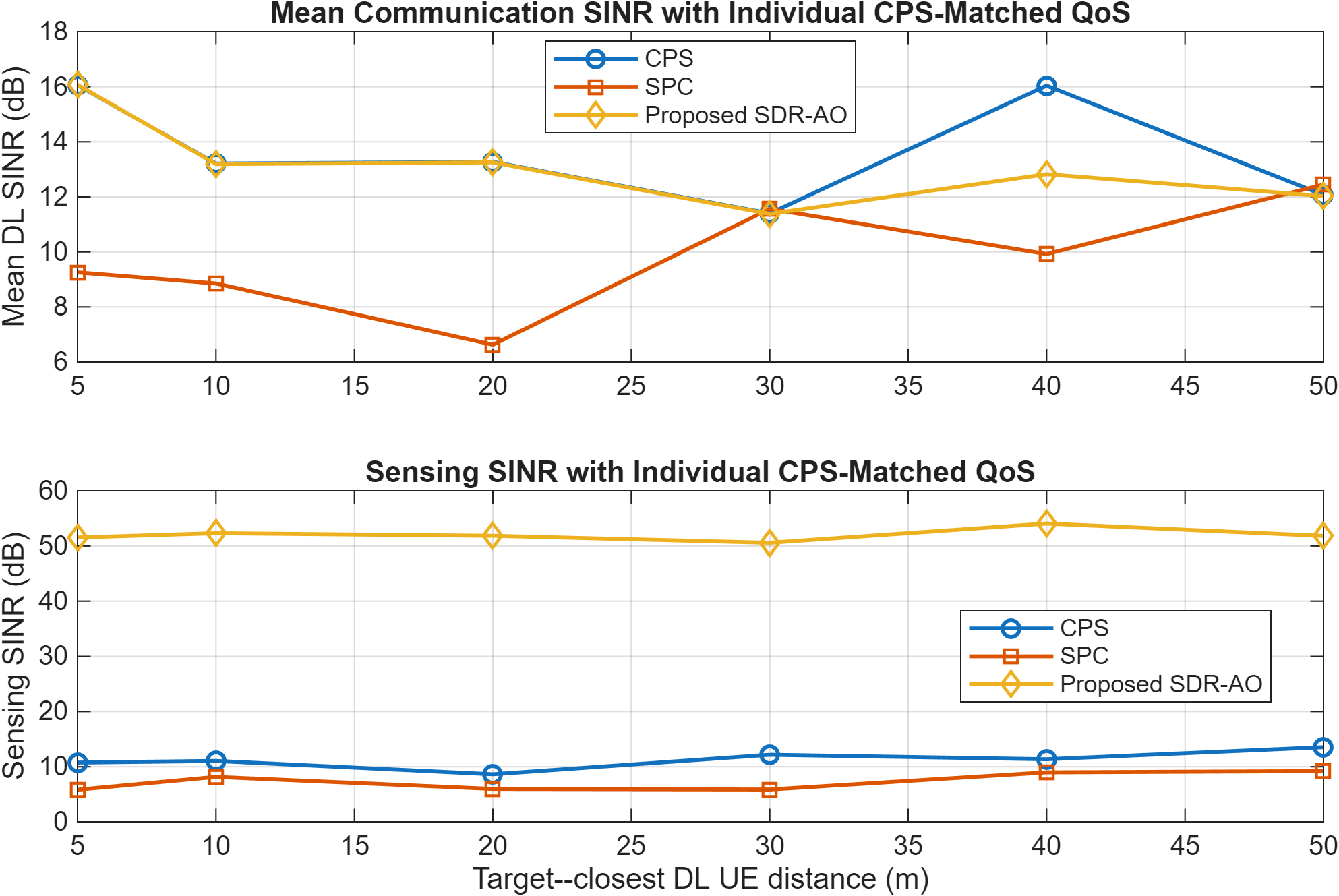}
	\caption{Communication and sensing performance versus the
		target--closest-DL-UE distance under individual CPS-matched QoS
		requirements.}
	\label{fig:individual_qos_comparison}
	\label{fig:comp1}
\end{figure}

To further evaluate the ability of the proposed design to balance sensing and
communication, Fig.~\ref{fig:comp1} considers individual
DL QoS requirements derived from the communication-prioritized sensing (CPS)
solution. Specifically, for each network geometry, the DL SINRs achieved by
CPS are used as reference communication requirements for the proposed
SDR-AO optimization. The target-to-closest-DL-UE distance is then varied from
$5$~m to $50$~m while the AP and UE locations are kept unchanged within each
realization. This experiment therefore examines whether the proposed method
can exploit the spatial degrees of freedom created by different target--UE
geometries while maintaining communication performance comparable to the
communication-prioritized reference.

The upper subplot of Fig.~\ref{fig:comp1} shows that the
proposed SDR-AO solution closely follows the CPS communication performance
over most of the considered distances. For example, at target--UE distances
of approximately $5$, $10$, $20$, and $30$~m, the mean DL SINRs obtained by
the proposed solution are nearly identical to those of CPS. At $40$~m, CPS
achieves a higher mean SINR, approximately $16$~dB compared with approximately
$13$~dB for the proposed design. This difference does not contradict the
optimization objective because the proposed method is not designed to
maximize the communication SINR beyond the imposed QoS requirements. Once
the required communication performance is satisfied, the remaining spatial
and power resources can instead be exploited to improve the sensing
objective. The SPC approach generally provides lower communication SINR,
particularly for small and intermediate target--UE separations, since sensing
is prioritized before the communication beamforming stage.

The benefit of this resource allocation is evident in the lower subplot.
The proposed SDR-AO scheme maintains a sensing SINR of approximately
$50$--$54$~dB throughout the complete target-distance range. By comparison,
CPS achieves approximately $8$--$13$~dB, while SPC provides approximately
$5$--$9$~dB. Thus, although the proposed solution is constrained to preserve
the individual communication QoS levels derived from the
communication-prioritized reference, it provides a very substantial sensing
gain.

Another important observation is that the sensing performance of the proposed
method remains relatively insensitive to the target--UE separation. The
sensing SINR varies only by a few decibels as the distance changes from
$5$~m to $50$~m. This indicates that the joint SDR-AO design can adapt the
communication and sensing covariance matrices to changes in the spatial
relationship between the sensing target and communication users. In contrast,
the priority-based schemes possess fewer degrees of freedom because one
function is designed first and the other is subsequently accommodated using
the remaining spatial or power resources.

Overall, Fig.~\ref{fig:comp1} demonstrates the central
advantage of the proposed formulation: the communication QoS constraints do
not require the optimizer to maximize communication SINR unnecessarily.
Instead, once the individual DL requirements are fulfilled, the optimization
can exploit the remaining degrees of freedom to maximize sensing performance.
This explains why the proposed SDR-AO solution can provide communication
performance close to the CPS reference while simultaneously achieving a
substantially larger sensing SINR.

\subsection{Convergence of the Proposed SDR--AO Algorithm}

\begin{figure}[t]
	\centering
	\includegraphics[width=\linewidth]{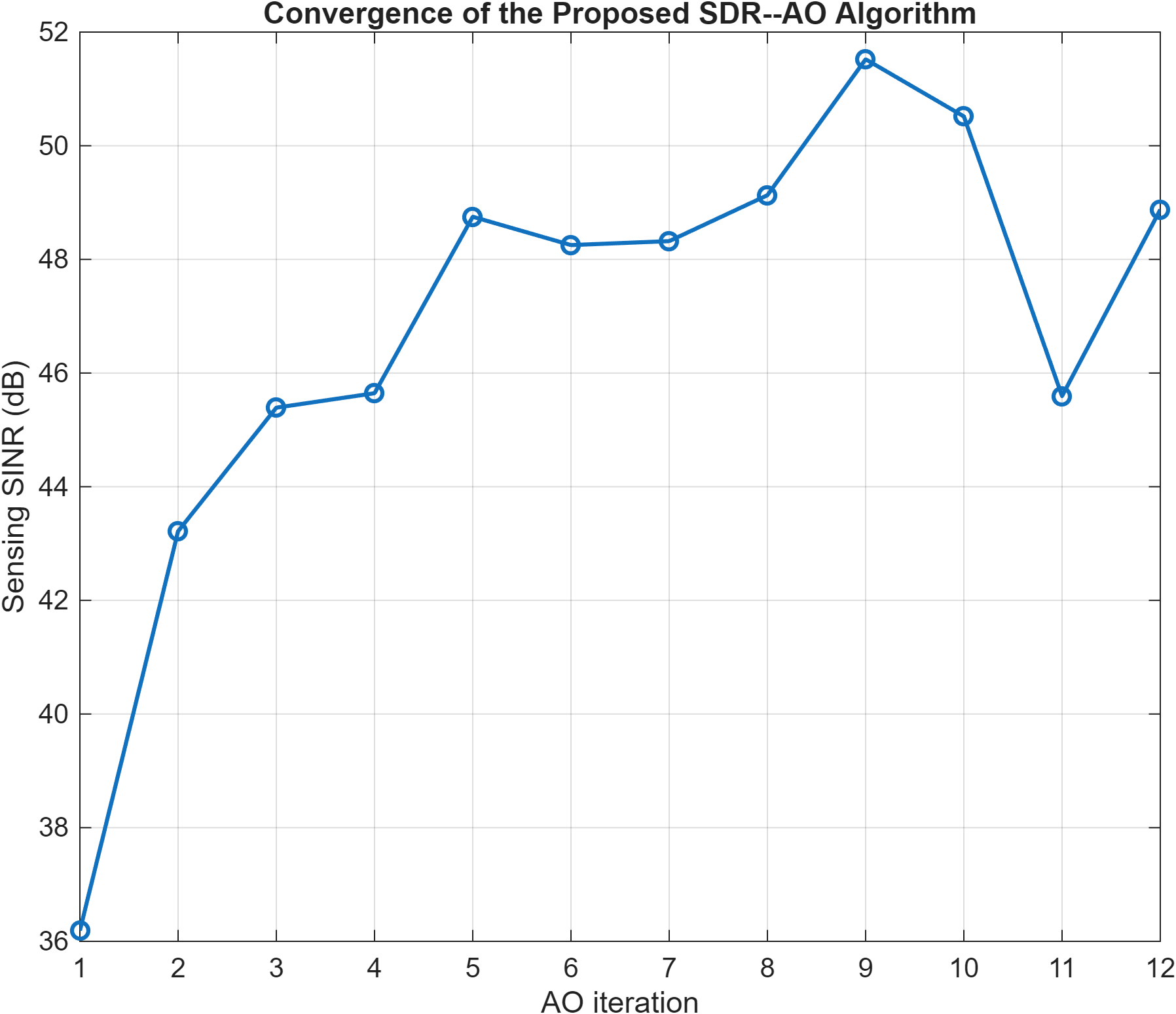}
	\caption{Evolution of the sensing SINR over the iterations of the proposed SDR--AO algorithm.}
	\label{fig:ao}
\end{figure}

Fig.~\ref{fig:ao} presents the evolution of the optimized sensing SINR over the AO iterations. The sensing SINR starts at approximately $36.1$~dB and rapidly increases to about $43.2$~dB after the second iteration. It further increases to approximately $48.7$~dB by the fifth iteration.

The sensing SINR subsequently varies within approximately $45.6$--$51.5$~dB. The highest value in the illustrated realization is obtained around the ninth iteration, where the sensing SINR reaches approximately $51.5$~dB. The final value after $12$ recorded iterations is approximately $48.9$~dB.

It is worth noting that the objective sequence is not strictly monotonic in this realization. This behavior results from alternating between the SDR beamforming update, UL power-allocation update, and the additional DL-SINR refinement step. Therefore, the value produced by one block is not necessarily preserved after the subsequent block update.

\subsection{Communication SINR Performance}

\begin{figure}[t]
	\centering
	\includegraphics[width=\linewidth]{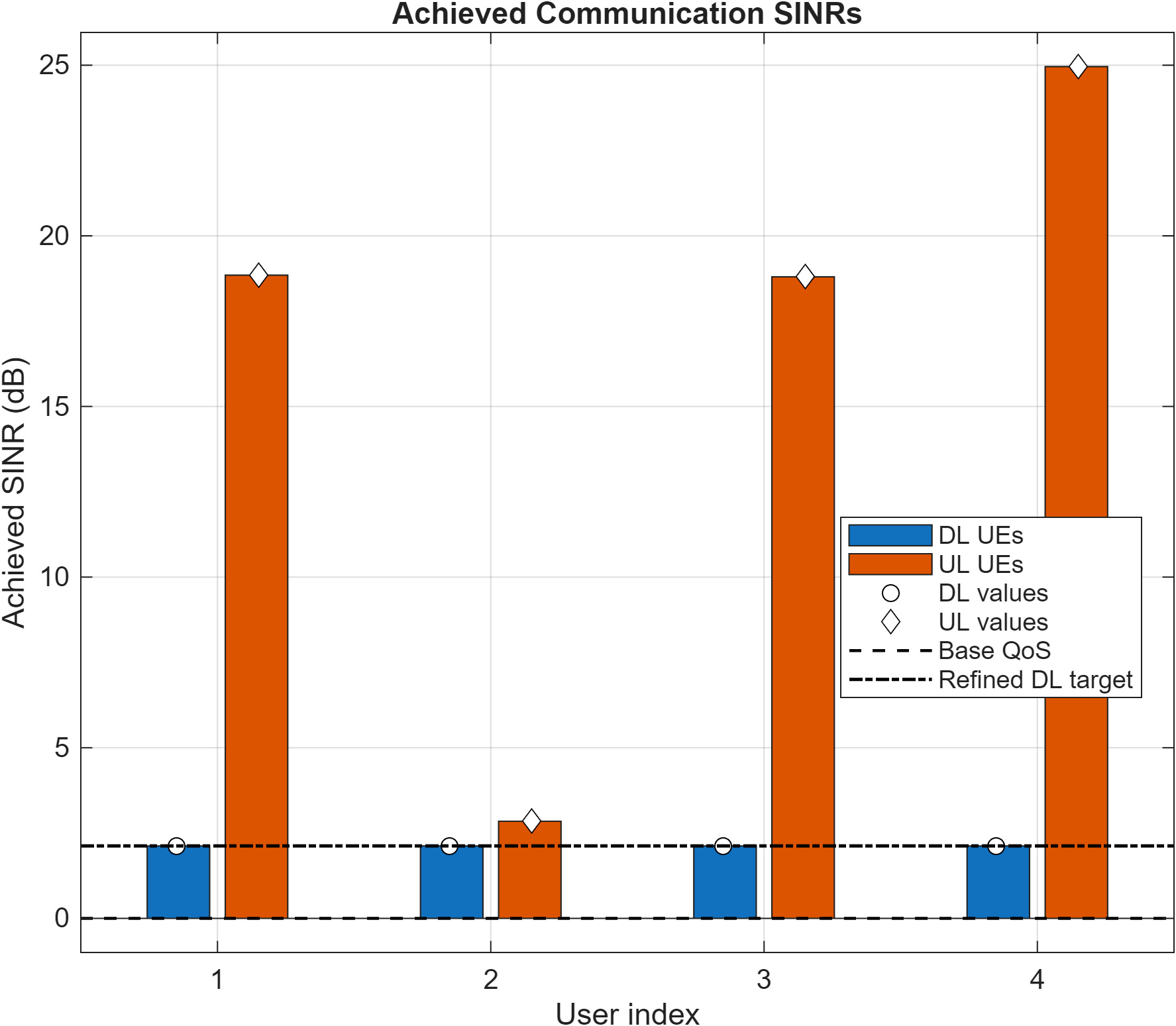}
	\caption{Achieved DL and UL communication SINRs for the considered network realization.}
	\label{fig:user_sinr}
\end{figure}

The achieved DL and UL communication SINRs are shown in Fig.~\ref{fig:user_sinr}. The four DL UEs achieve approximately the same SINR of $2.1$~dB. Hence, all DL users satisfy both the original $0$~dB QoS requirement and the refined common DL target.

The achieved UL SINRs are considerably different among users. In the illustrated realization, the UL SINRs are approximately $18.9$, $2.8$, $18.8$, and $25.0$~dB for UL UEs $1$--$4$, respectively. All UL UEs therefore satisfy the original $0$~dB minimum SINR requirement.

The nearly identical DL SINRs are a consequence of the secondary max--min DL refinement. In contrast, the UL powers are optimized primarily to satisfy the communication constraints while controlling their contribution to the sensing interference. Consequently, the UL SINRs are not explicitly forced toward a common value.

\subsection{UL Power-Allocation Evolution}

\begin{figure}[t]
	\centering
	\includegraphics[width=\linewidth]{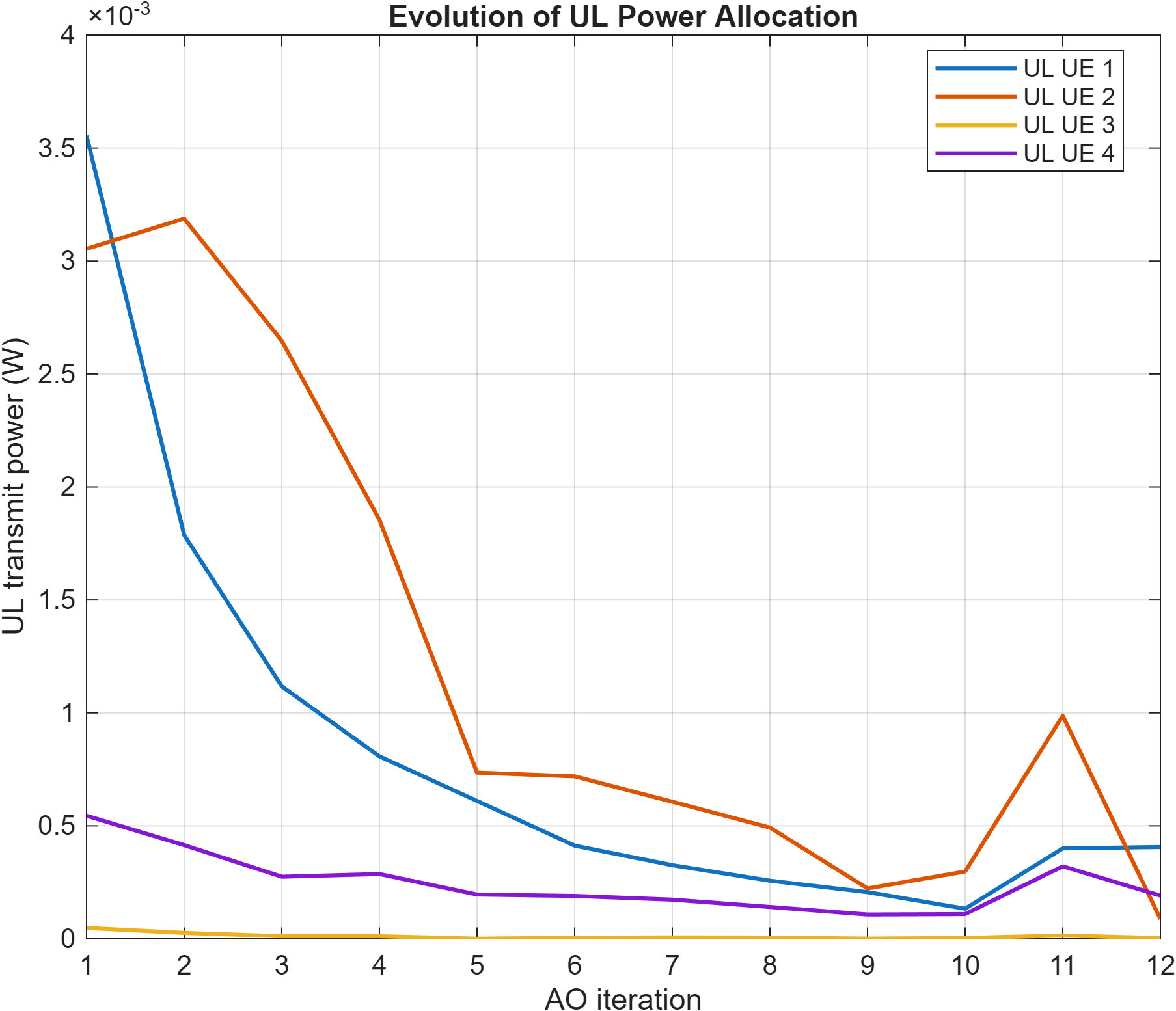}
	\caption{Evolution of the optimized UL UE transmit powers over the AO iterations.}
	\label{fig:ul_power}
\end{figure}

Fig.~\ref{fig:ul_power} illustrates the UL transmit-power evolution during the AO procedure. All optimized UL powers remain significantly below the maximum available UL transmit power of $0.1$~W.

For example, the power of UL UE~1 decreases from approximately $3.5\times10^{-3}$~W during the first iteration to approximately $4\times10^{-4}$~W at the final iteration. The power allocated to UL UE~2 also decreases considerably over the AO iterations, although a temporary increase is observed at iteration $11$. UL UE~3 requires only a very small transmit power throughout the optimization, whereas UL UE~4 operates at an intermediate power level.

These results demonstrate that the UL communication constraints can be satisfied without requiring the UEs to operate close to their maximum available transmit powers for the considered channel realization.

\subsection{Per-AP Transmit-Power Utilization}

\begin{figure}[t]
	\centering
	\includegraphics[width=\linewidth]{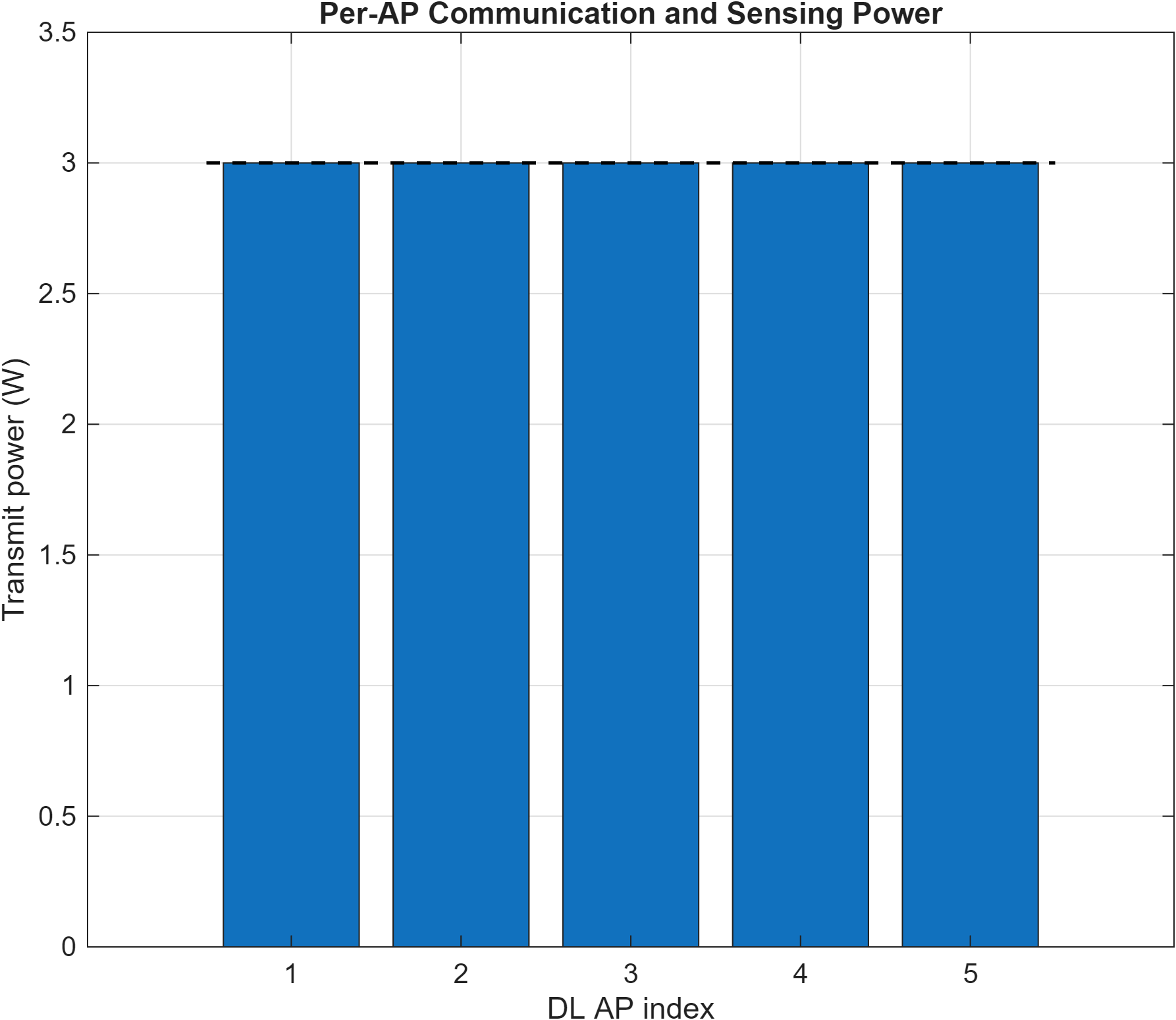}
	\caption{Optimized total transmit power of the DL APs and the corresponding per-AP power constraint.}
	\label{fig:ap_power}
\end{figure}

Fig.~\ref{fig:ap_power} shows the total optimized transmit power of each DL AP. All five DL APs operate approximately at their maximum allowed power,
\begin{equation}
	P_m^{\mathrm{DL}}\simeq3~\mathrm{W},
	\qquad
	m=1,\ldots,5.
\end{equation}
The dashed horizontal line in Fig.~\ref{fig:ap_power} denotes the $3$~W per-AP power constraint.

The active power constraints are consistent with the sensing-SINR maximization objective, since additional feasible DL transmit power can contribute to target illumination while the communication and interference constraints remain satisfied.

\subsection{MAPRT Target-Detection Performance}

\begin{figure}[t]
	\centering
	\includegraphics[width=\linewidth]{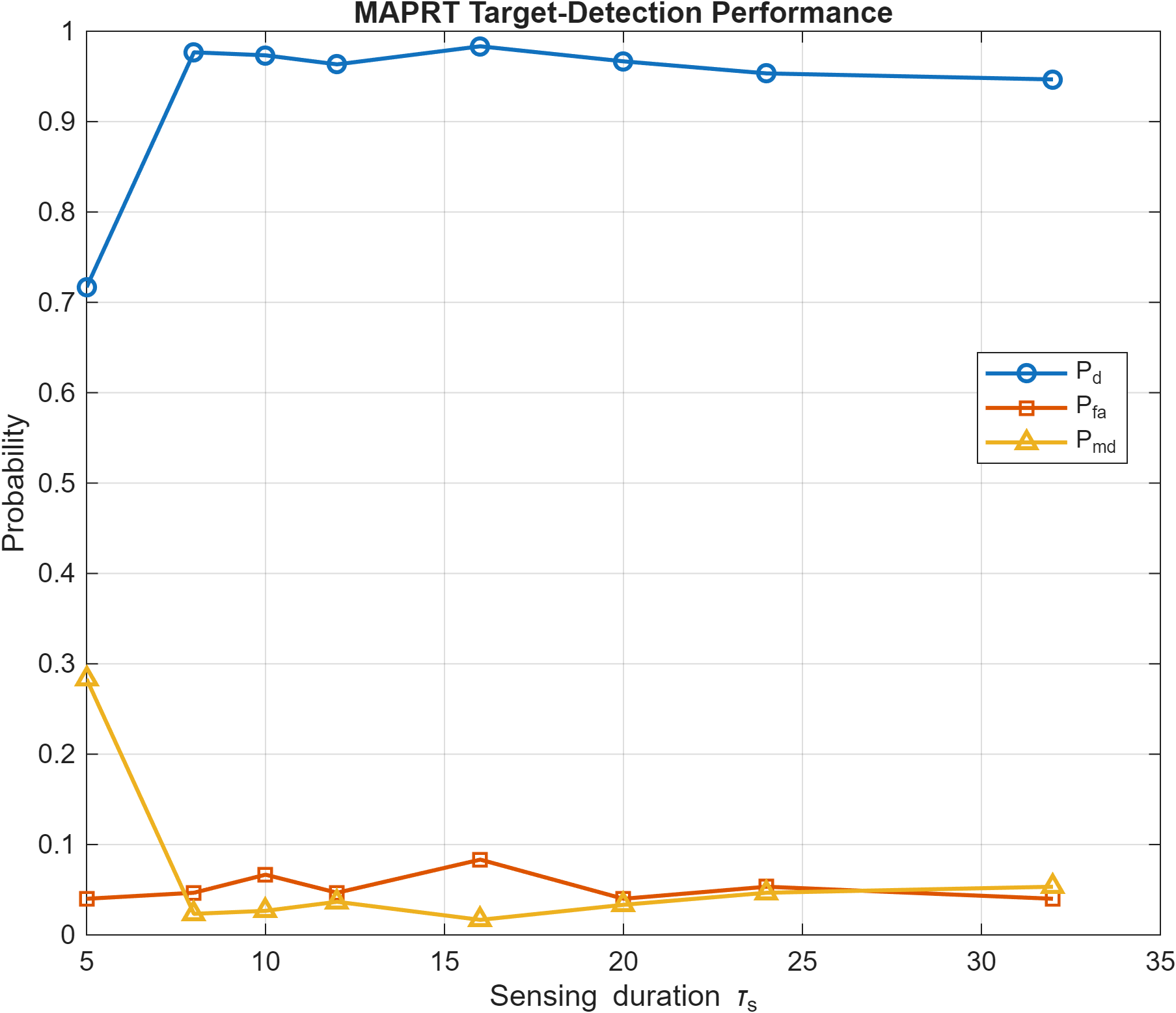}
	\caption{MAPRT target-detection performance versus the sensing duration $\tau_s$.}
	\label{fig:detection}
\end{figure}

Fig.~\ref{fig:detection} presents the target-detection performance of the MAPRT detector as a function of the sensing duration $\tau_s$. The three quantities shown are the probability of detection $P_d$, probability of false alarm $P_{fa}$, and probability of misdetection
\begin{equation}
	P_{md}=1-P_d.
\end{equation}

At $\tau_s=5$, the obtained detection probability is approximately $0.71$, corresponding to a misdetection probability of approximately $0.29$. Increasing the sensing duration to $\tau_s=8$ significantly improves the detection probability to approximately $0.98$.

For $\tau_s\geq8$, the detector maintains a high detection probability. The obtained values of $P_d$ remain approximately within the range $0.95$--$0.98$, while $P_{md}$ remains approximately within $0.02$--$0.05$. At the default value $\tau_s=10$, the detection probability is approximately $0.97$.

The empirical false-alarm probability fluctuates around the desired value of $0.05$. Across the considered sensing durations, $P_{fa}$ lies approximately between $0.04$ and $0.08$, with the observed variation resulting partly from the finite number of Monte Carlo realizations.

The results indicate that the largest detection improvement occurs when increasing $\tau_s$ from $5$ to $8$. Beyond this point, increasing the sensing duration provides relatively limited additional detection improvement for the considered simulation setting.

\subsection{Separation of the MAPRT Test Statistic}

\begin{figure}[t]
	\centering
	\includegraphics[width=\linewidth]{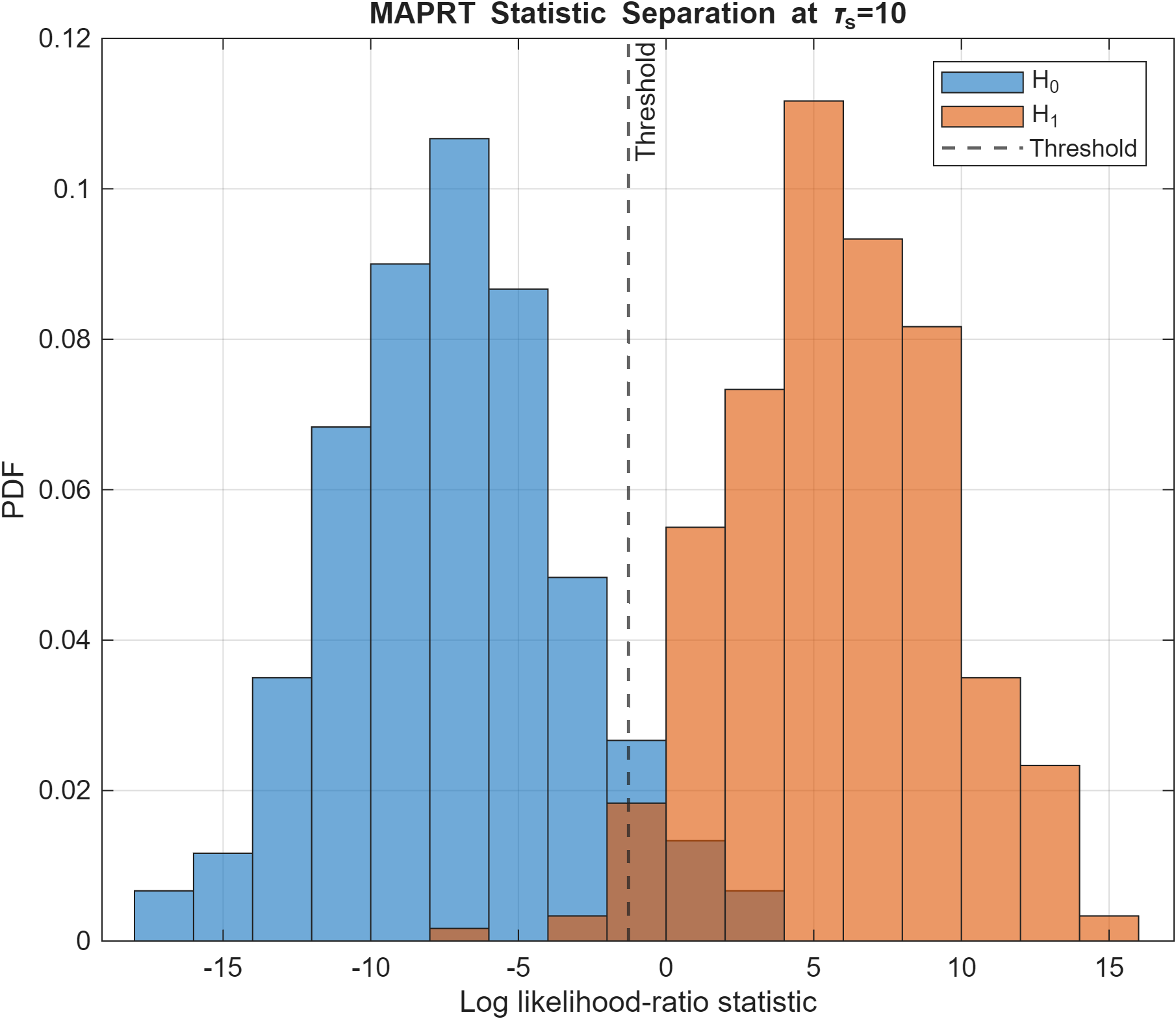}
	\caption{Empirical distributions of the MAPRT log-likelihood-ratio statistic under $\mathcal{H}_0$ and $\mathcal{H}_1$ for $\tau_s=10$.}
	\label{fig:maprt_statistics}
\end{figure}

To further examine the behavior of the detector, Fig.~\ref{fig:maprt_statistics} presents the empirical distributions of the log-likelihood-ratio statistic under hypotheses $\mathcal{H}_0$ and $\mathcal{H}_1$ at $\tau_s=10$.

Under $\mathcal{H}_0$, the MAPRT statistic is concentrated mainly at negative values, with its distribution centered approximately around $-6$. In contrast, under $\mathcal{H}_1$, the statistic is shifted considerably toward positive values and is concentrated mainly around $4$--$8$. The calibrated decision threshold is located approximately at $-1.3$.

Only a relatively small overlap exists between the two distributions around the decision boundary. This separation is consistent with the high value of $P_d$ and the relatively small $P_{fa}$ obtained at $\tau_s=10$ in Fig.~\ref{fig:detection}.

\subsection{Optimized Sensing Beampattern}

\begin{figure}[t]
	\centering
	\includegraphics[width=\linewidth]{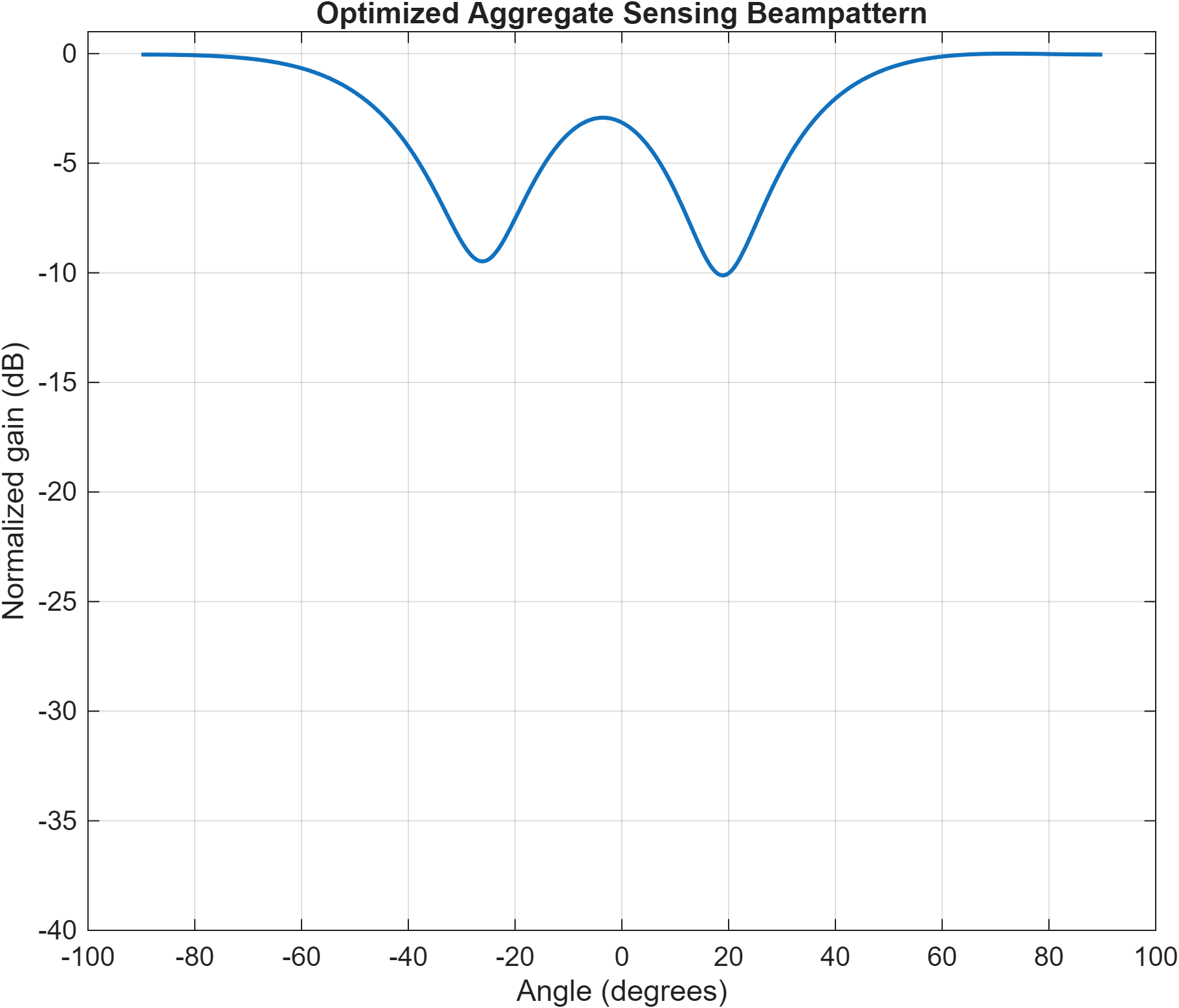}
	\caption{Normalized aggregate sensing beampattern corresponding to the optimized transmit covariance matrices.}
	\label{fig:beampattern}
\end{figure}

Fig.~\ref{fig:beampattern} shows the normalized aggregate sensing beampattern obtained from the optimized transmit covariance matrices. The pattern exhibits two pronounced attenuation regions around approximately $-26^\circ$ and $19^\circ$, where the normalized gains decrease to approximately $-9.5$ and $-10$~dB, respectively.

The normalized gain approaches $0$~dB toward the outer angular regions, while a local level of approximately $-3$~dB is observed around broadside.

Since the considered cell-free architecture consists of geographically distributed APs rather than a single co-located antenna array, Fig.~\ref{fig:beampattern} should be interpreted as an aggregate angular representation of the optimized covariance solution rather than the conventional beampattern of a single ULA.
```
%=========================================================
% CONCLUSION
%=========================================================

\section{Conclusion} \label{sec:conclusion}This paper studied priority-based beamforming for an NAFD cell-free ISAC network supporting simultaneous DL communication, UL communication, and distributed target sensing. Channel estimation and MAPRT-based target detection were explicitly incorporated into the transmission protocol before target-aware beamforming was performed. Two complementary beamforming strategies were considered. In the communication-prioritized sensing design, the sensing beam was constructed in the nullspace of the effective communication channels, thereby protecting the DL UEs and UL receivers from additional sensing-induced interference. In the sensing-prioritized communication design, the sensing beam was selected first according to the target direction, and the communication beamformers were then optimized using a max--min SINR formulation under the NAFD interference and per-AP power constraints. These two schemes provide useful benchmark operating points for studying the sensing--communication tradeoff. The numerical results showed that reliable target detection and communication QoS can be simultaneously maintained in the considered NAFD cell-free deployment. In particular, the MAPRT detector achieved approximately $P_d=0.97$ at $\tau_s=10$ with a false-alarm probability close to $0.05$. The results also illustrate how the choice of priority directly affects the available spatial degrees of freedom, communication interference, and target illumination. Future work may extend these priority-based designs to dynamic AP-mode selection, multi-target sensing, robust beamforming under channel and target-location uncertainty, and adaptive selection of the communication--sensing priority according to network conditions
%=========================================================
% REFERENCES
%=========================================================

\end{document}